\documentclass[prd,preprint,aps,superscriptaddress,showpacs,preprintnumbers,nofootinbib]{revtex4-1}
\pdfoutput=1 
\usepackage{varioref,exscale,latexsym,amsmath,amssymb}
\usepackage[pdftex]{graphicx}
\usepackage{hyperref}
\usepackage{bm}
\usepackage{slashed}
\usepackage{epsfig}
\usepackage{color}
\usepackage[utf8]{inputenc}
\usepackage{feynmp}
\usepackage{subcaption}
\usepackage[utf8]{inputenc}
\usepackage[normalem]{ulem}
\usepackage{color}  
\allowdisplaybreaks

\newcommand{\be}{\begin{equation}} 
\newcommand{\ee}{\end{equation}}  
\newcommand{\bea}{\begin{eqnarray}}  
\newcommand{\eea}{\end{eqnarray}}

\newcommand{\hl}{}

\makeatletter
\newcommand*{\rom}[1]{\expandafter\@slowromancap\romannumeral #1@}
\makeatother
\begin{document}
\title{
Boson Star from Repulsive Light Scalars and Gravitational Waves
}

\author{Djuna Croon}\email{dcroon@triumf.ca}
\affiliation{TRIUMF Theory Group, 4004 Wesbrook Mall, Vancouver, B.C. V6T2A3, Canada}

\author{JiJi Fan}\email{jiji$\_$fan@brown.edu}
\affiliation{Department of Physics, Brown University, Providence, RI, 02912, USA}

\author{Chen Sun}\email{chen.sun@brown.edu}
\affiliation{CAS Key Laboratory of Theoretical Physics, Institute of
  Theoretical Physics, Chinese Academy of Sciences, Beijing 100190,
  P. R. China}
\affiliation{Department of Physics, Brown University, Providence, RI, 02912, USA}

\date{\today}

\begin{abstract}
We study properties of boson stars consisting of ultra-light scalar dark matter with repulsive self-interactions. We investigate the origin of the maximum mass of spherically symmetric stable stars which emerges only when solving the full equations of motion in curved space-time, but not when solving the approximated Schr\"odinger-Newton equations. When the repulsion is weak, the backreaction of the curvature on the scalars acts as an additional source of attraction and can overcome the repulsion, resulting in a maximum star mass and compactness. We also point out that the potential in a UV completed particle physics model of light scalar dark matter is generally more complicated than the widely used $\phi^4$ interaction. Additional interactions beyond $\phi^4$ in the potential can dramatically change the properties of boson stars as well as modify the prospect of LIGO gravitational wave detection for binary mergers of boson stars.

\end{abstract}

\pacs{xxx.xxx}

\preprint{xxx-xxx}

\maketitle

\tableofcontents

\section{Introduction}

The detection of gravitational waves (GW) from black hole (BH) mergers \cite{TheLIGOScientific:2016agk,TheLIGOScientific:2016qqj} and neutron star (NS) merger \cite{GBM:2017lvd} indicates the advent of an era of GW astronomy. A natural question one can ask is whether and how this GW probe is sensitive to new physics beyond the Standard Model (SM).
As the merger events detected are astrophysical, the signal is of a classical nature and seems not directly sensitive to sub-atomic quantum physics. Yet if the microphysics beyond the SM leads to interesting predictions on macroscopic scales, i.e., of the size of BH and NS, GW probes may be relevant. 

Among many different proposals to use GW to probe new physics (e.g., \cite{Giudice:2016zpa, Dev:2016hxv, Ellis:2017jgp,Croon:2017zcu,Randall:2017jop, Cui:2017ufi, Pierce:2018xmy, Amin:2018kkg, Nelson:2018xtr, Geller:2018mwu, Croon:2018erz,Figueroa:2018xtu, Bauswein:2018bma,Kopp:2018jom,Alexander:2018qzg}), we focus on the scenario with a very light scalar as a dark matter candidate. 
It has long been conjectured that very light scalars with long de Broglie wavelength and suppressed couplings to the SM may form Bose-Einstein condensates (BEC), which can lead to macroscopically sized clumps called boson stars~\cite{Tkachev:1986tr}. As every scalar particle is in its ground state, the star can be described effectively by a single wave function. Therefore, the boson star properties can be inferred from the underlying microscopic physics, which makes it a good candidate to bridge the gap between microscopic physics and the GW signal. In particular, we will consider mergers of two boson stars, being agnostic of their formation history. 

The scalar bosons that may give rise to boson stars fall into two main categories, depending on the sign of the bosons' leading self-interactions in the non-relativistic (NR) limit. If the coefficient of the leading interaction term in the potential is negative, the scalar bosons have attractive self-interactions; otherwise, they interact repulsively. A classic example of bosons with attractive self-interactions is the QCD axion~\cite{Peccei:1977ur, Peccei:1977hh, Wilczek:1977pj,Weinberg:1977ma, Kim:1979if, Shifman:1979if, Zhitnitsky:1980tq,Dine:1981rt}. It has been demonstrated recently that the only cosmologically stable QCD axion star is a dilute star~\cite{Visinelli:2017ooc, Schiappacasse:2017ham, Chavanis:2017loo, Eby:2017teq}, which results from a balance between gravity and repulsive kinetic pressure. For the most quoted benchmark with axion decay constant at $6 \times 10^{11}$ GeV and axion mass at $10^{-5}$ eV, the maximally dense stable axion star mass is $\sim 10^{-11}$M$_{\odot}$, far below the sensitivity of GW detectors - even if two such objects would merge. On the other hand, light scalars with repulsive self-interactions can potentially form much heavier stars since their self-interactions provide an additional source to balance against gravity, which tends to shrink and collapse the star. In this note, we will focus on the dark stars formed of light scalars with repulsive self-interactions and their possible GW signals.\footnote{It has been suggested (e.g., \cite{Goodman:2000tg,Peebles:2000yy}) that repulsive interaction may help alleviate core-cusp problem~\cite{Flores:1994gz, 1994Natur.370..629M, Burkert:1995yz, Moore:1999gc, Salucci:2000ps}. Yet~\cite{Deng:2018jjz} argues that light scalar dark matter, whatever their self-interactions are, cannot address the problem in general. }

The spherically symmetric boson star solution with repulsive self-interaction has been found more than 30 years ago in \cite{Colpi:1986ye}. There have been intensive studies on boson
stars. A nice and highly cited review on this topic can be found in~\cite{Schunck:2003kk}. An early review can be found in \cite{Jetzer:1991jr}. After LIGO, possible observable GW signals based on this classic solution have been studied in several papers, e.g.~\cite{Giudice:2016zpa,Croon:2018ftb, Helfer:2018vtq, Widdicombe:2018oeo, Bezares:2018qwa}. We will revisit the computations and make two new observations, which are summarized below: 
\begin{itemize}
\item {\hl We explain, {\hl from the force balancing point of view,}
    why there is no stable boson star solution beyond a certain
    compactness in the phenomenological scalar model with a repulsive
    $\left|\phi\right|^4$ interaction, as first observed in
    \cite{Colpi:1986ye}.} Working in the NR and flat space limits and
  solving the Schr\"odinger-Newton equations of motion, one would find
  that a repulsive $\left|\phi\right|^4$ interaction leads to an
  indefinite growth of the boson star, as demonstrated
  in~\cite{Chavanis:2011zi,Chavanis:2011zm,Schiappacasse:2017ham}. Yet
  if one makes no approximation and solves the Einstein-Klein-Gordon
  equations as in \cite{Colpi:1986ye} {\hl and more recently in
    \cite{Chavanis:2011cz}}, there is no stable solution beyond a
  certain compactness. {\hl We find that} the instability of such a
  system at large compactness may be due to an additional source of
  attraction from the back-reaction of the space-time curvature on the
  boson stars, which grows as $N^2$ (the same proportionality as the
  repulsion and gravity) with $N$ being the particle number of the
  star. This is quite different from the QCD axion star case with
  attractive self-interaction, where the maximum mass is achieved at
  the point where self-interaction becomes important. {\hl Alternatively,
    the dynamics of such a system can be studied and the maximal mass can be understood using continuity and hydrodynamic Euler equations as demonstrated in
    \cite{Chavanis:2011cz}.}  

Yet on the other hand, numerical computations in \cite{Colpi:1986ye} were carried out for relatively weak repulsion: identifying the light scalar dark matter as a pseudo-Nambu-Goldstone boson (pNGB), the corresponding decay constant is then $\gtrsim 10^{16}$ GeV (the larger the decay constant is, the smaller the repulsion is). It is currently not understood what happens when we increase the repulsion or equivalently lower the decay constant to be below $\lesssim 10^{16}$ GeV. 

\item
We point out that taking the underlying particle physics explanation for a light scalar into account, the boson stars' properties can be dramatically different from those derived in \cite{Colpi:1986ye}. 
In \cite{Colpi:1986ye}, the scalar potential is approximated by a single $\left|\phi\right|^4$ interaction. Yet if one tries to explain the lightness of the scalar in a more complete particle physics model, for example, by identifying the light scalar as a pNGB so that its mass is protected by an approximate shift symmetry, one always finds that the scalar potential is significantly more complicated with a series of higher-order terms~\cite{Fan:2016rda}. These higher-order terms can become as important as the leading quartic interaction in a more heavy and dense star, modifying the prediction of maximum mass and compactness. {\hl They will also change the prospect of GW detection for merging of two boson stars, if the boson stars can be formed and are stable on cosmological time scale.}
\end{itemize}

{\hl The rest of this article is organized as follows. In Section~\ref{sec:basics-dark-star}, we introduce the effective Lagrangian of light scalar dark matter models and discuss the solutions and their stability properties in Schr\"odinger-Newton picture, both numerically and analytically. We also review some basic estimates of gravitational waves from a general binary merger system. In Section~\ref{sec:re-visit-space}, we solve the full general relativity (GR) system with a repulsive $\lambda |\phi|^4$ interaction, again both numerically and analytically. We use a set of ansatz to show that the maximal star mass and compactness in the repulsive theory originates from balance between the back-reaction of the space-time curvature and repulsion. In Section~\ref{sec:realistic-dark-star}, we introduce a natural light scalar dark matter model with repulsive self-interactions beyond the $|\phi|^4$. We find the star solutions and compute the corresponding maximum mass and compactness to demonstrate that they are dramatically different from those from a simple $|\phi|^4$ interaction. We also present the parameter space which leads to detectable mergers of boson stars by LIGO. We conclude and discuss possible future directions in Section~\ref{sec:conclusion}. }

\section{Basics of Light Scalar Dark Matter, 
  Dark Stars, and Gravitational Waves}
\label{sec:basics-dark-star}
\subsection{Effective Lagrangian of pNGB}
Let us start with an effective Lagrangian of a light complex scalar, $\phi$, with interaction
up to $|\phi|^4$.
 \be
  \label{eq:model}
  \mathcal{L}
=
\frac{1}{2}\left|\partial_\mu \phi \right|^2 -
\frac{1}{2}m^2| \phi|^2 -
    \frac{\lambda}{4} \frac{m^2}{f^2}|\phi|^4 + \cdots,
\ee
where  $\lambda$ is a dimensionless quantity (with natural values of order one) and $f$ is of mass dimension one. The dots represent high dimension operators, suppressed by more powers of $f$. This parametrization is motivated by identifying the scalar dark matter as a pNGB {\hl so that its lightness is protected by an approximate shift symmetry.} Then $f$ is usually referred to as the decay constant and {\hl is the symmetry breaking scale}. In general, a pNGB can be real or complex, depending on the symmetry breaking pattern.\footnote{If the pNGB is a real scalar, there could be additional terms in the Lagrangian such as $m^2\phi^3/f$.} {\hl We want to emphasize that whether the pNGB is real or complex}, the form of the effective Lagrangian with $m^2$ being the symmetry breaking spurion and appearing in front of the interaction terms stay the same. 
 
In the NR limit, to leading order in $\mathbf{p'}/m$, we can write {\hl a positive energy eigenstate in position space as
\begin{align}
  \label{eq:real-scalar}
  \phi(x^\mu) 
  =
    \frac{1}{\sqrt{m}} 
    \psi(\mathbf{x},t) \mathrm{e}^{-imt},
\end{align}}
where the complex field $\psi(\mathbf{x},t)$ is of mass dimension
$3/2$, and the time scale of its variation is $\gg 1/m$. 
The time dependence of the oscillation modes is dominantly controlled
by the $e^{-imt}$ term. Plugging Eq.~\eqref{eq:real-scalar} into
Eq.~\eqref{eq:model} and discarding fast oscillating terms with
phase $\propto mt$ (in the NR limit, these terms average to be zero
over the physical time scale, which is $\gg 1/m$), we obtain the
leading NR Lagrangian of the theory as follows: 
\begin{eqnarray}
  \label{eq:KG-with-int-NR}
  \mathcal{L}
  & \approx &
    \frac{i}{2}
    (\dot \psi \psi^* - \dot \psi^* \psi)
    -    \frac{1}{2m}
    | \nabla \psi|^2 
    - \frac{\lambda}{4f^2} |\psi|^4, 
\end{eqnarray}
where the dot means a derivative with respect to time and we ignore $|\dot{\psi}|^2$ and higher order terms. Without gravity, the equation of motion (EOM) is then given by
\begin{eqnarray}
  \label{eq:NR-limit-EOM}
    i\dot \psi
  & =&
    -
    \frac{1}{2m} \nabla^2 \psi + \frac{{\lambda}}{2 f^2}
    |\psi|^2 \psi,
\end{eqnarray}
In the spherically symmetric case, the ground state wave function $\psi(r)$ decreases monotonically, \textit{i.e.},
$\partial |\psi|/\partial r <0$. In the classical limit, the force
between the scalars is given by
$F_r = - \partial V/ \partial r \propto - (\partial |\psi|/\partial r)
{\lambda} $.  Therefore, we refer to
${\lambda}>0 (<0)$ as the repulsive (attractive) interaction,
respectively.

\subsection{Dark Star Solutions in Flat Space-Time}
\label{sec:dark-star-solutions}
In this section, we {\hl briefly review the solutions of dark star in the NR and flat space limits.} We also examine the stability of the solutions. Some discussions follow Refs.~\cite{Schiappacasse:2017ham} closely, yet with a different focus.
In the NR and flat space limits, the EOM for the scalar field, $\psi$, defined in Eq.~\ref{eq:NR-limit-EOM}, at the leading order, is the Schr\"odinger-Newton (or also known as the Schr\"odinger-Poisson) equation, 
\begin{equation}
    i\dot \psi =
    -
    \frac{1}{2m} \nabla^2 \psi + \frac{{\lambda} }{2 f^2} \;
    |\psi|^2\psi
            -G_Nm^2\psi  \int d^3 \mathbf{x'} \frac{\psi^*(\mathbf{x'})
    \psi(\mathbf{x'})}{|\mathbf{x} - \mathbf{x'}|},
      \label{eq:schrodinger-newton}
\end{equation}
where gravity is taken into account in the last term. 
The energy of such a system can be broken into kinetic, self-interaction, and gravitational energies, 
\begin{equation}
  H = H_{kin} + H_{int} + H_{grav}, 
    \label{eq:energy-of-NR-system}
\end{equation}
where 
\begin{eqnarray}
  \label{eq:NR-system-energy-components}
  H_{kin}
  &=&
    \frac{1}{2m} \int d^3\mathbf{x} \; \nabla \psi^*(\mathbf{x}) \nabla \psi(\mathbf{x}),
    \cr
    H_{int}
  & =&
    \frac{{\lambda} }{4 f^2}    \int d^3\mathbf{ x} \; |\psi(\mathbf{x})|^4,
    \cr
    H_{grav}
  & =&
    - \frac{G_Nm^2}{2}    \int d^3 \mathbf{x}\; \psi (\mathbf{x})\psi^* (\mathbf{x}) \int d^3 \mathbf{x'} \frac{\psi^*(\mathbf{x'})
    \psi(\mathbf{x'})}{|\mathbf{x} - \mathbf{x'}|}.
\end{eqnarray}
For a spherically symmetric dark star solution, it is convenient to
define the $n\%$ mass radius, denoted as $R_n$, which is the radius that encloses $n\%$ of the total star mass.
We choose $R_{90}$ to be the characteristic radius of such systems. A boson star solution satisfies $\partial H /\partial R_{90} = 0$. The solution is stable (unstable) if
$\partial^2 H/ \partial R^2_{90}>0 (<0)$.  Note that the stability and
even the existence of a boson star solution is not guaranteed by any
fundamental symmetry, but results from competition between different forces, i.e., the kinetic pressure due to uncertainty principle and gravity. We will elaborate more on this below.

\begin{figure}[h]
  \centering
  \includegraphics[width=.5\textwidth]{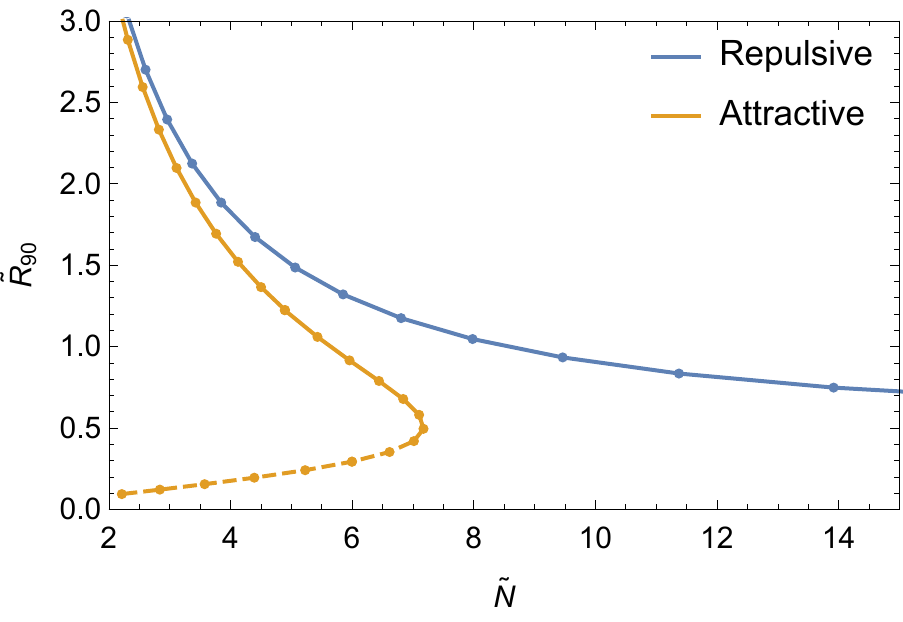
  }
  \caption{Numerical solutions of boson stars (each individual
    solution we find is represented by a dot) from the
    Schr\"odinger-Newton equation for attractive (orange) and
    repulsive (blue) interactions. We fix the parameters to be the
    same except for the sign of the quartic interaction. {\hl The dimensionless particle numbers and star radius, $\tilde{N}, \tilde{R}_{90}$, are defined in
    Eq.~\eqref{eq:dim-variables}. We fix $\lambda=-1$.}  }
  \label{fig:SN-attr-vs-repul}
\end{figure}

We seek stationary solutions using the single harmonic ansatz 
\begin{equation}
  \label{eq:harm-ansatz}
\psi(\mathbf{x},t) = e^{-i  \mu t} \psi(\mathbf{x}),
\end{equation}
and introduce the following dimensionless variables, which are denoted with a tilde on top
\begin{align}
\begin{array}{rlrlrl}
  \label{eq:dim-variables}
  \mathbf{x}
  & = \dfrac{\tilde{\mathbf{x}}}{\sqrt{G_N} fm},
  &N
  &=
    \tilde{N} \;    \dfrac{f}{\sqrt{G_N}m^2 } ,
  &    \psi
  &   = \tilde{\psi} \; \sqrt{G_Nm} f^2,
  \\
     \mu
  &=
           \tilde{\mu}\; G_Nf^2 m,
  &       H
  &= \tilde{H} \; \dfrac{\sqrt{G_N} f^3}{m}, 
  &\phi_N
  &= \tilde \phi_N \; G_Nf^2,
\end{array}
\end{align}
where $N = 4\pi \int dr r^2  \psi(r)^2$ is the total number of scalars
in the star, and $\phi_N$ is the Newtonian gravitational potential. In
this notation, the compactness of the star is expressed as
\begin{equation}
  \label{eq:compactness-dimless}
    C =
  \frac{G_NmN}{R}
      =
      \frac{\tilde N}{\tilde R}
      G_Nf^2,
\end{equation}
The
Schr\"odinger-Newton equation then can be rewritten in terms of these dimensionless quantities as 
\bea  \label{eq:EOM-with-correction-dimless}
  \tilde \mu \tilde \psi
  & =&
    -
    \frac{1}{2} \tilde  \nabla^2 \tilde \psi + \frac{{\lambda}}{2}\;
    |\tilde \psi|^2\tilde \psi
    +\tilde \psi \tilde \phi_N
        ,\cr
        \tilde \nabla^2 \tilde \phi_N
  & =&  4\pi |\tilde \psi|^2.
\eea
We solve the equations for both attractive ($\lambda < 0$) and
repulsive ($\lambda >$ 0) self-interactions and present the solutions
we find in the rescaled ($\tilde{N}, \tilde{R}_{90}$) plane in
Fig.~\ref{fig:SN-attr-vs-repul}.

From Fig.~\ref{fig:SN-attr-vs-repul}, one can see that for the case of attractive self-interactions, there is a maximum particle number that boson stars can reach. The lower branch beneath the turning point is unstable, {\hl which we represent in dashed curve.} On the other hand, in the repulsive case, the particle number continues to grow \cite{Chavanis:2011zi,Chavanis:2011zm,Schiappacasse:2017ham}. This result seems inconsistent with the earlier numerical results of \cite{Colpi:1986ye} with a turning point and {\hl an unstable branch} for the repulsive case as well. We will explain the origin of the discrepancy in Section \ref{sec:re-visit-space}. But before that, we discuss the stabilities of the solutions in the following subsection.

\begin{figure}[t]
  \centering
  \includegraphics[width=.5\textwidth]{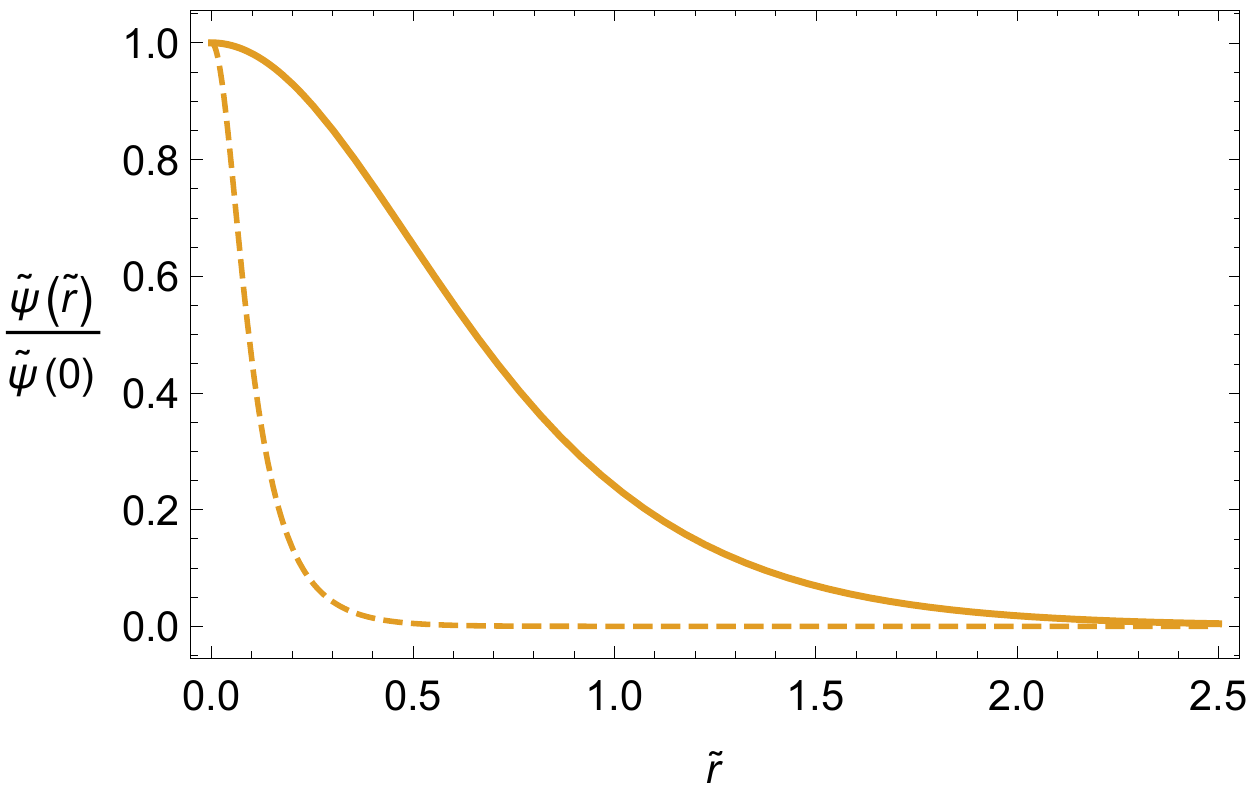
  }
  \caption{The shape of different possible wave functions at fixed $N$. The solid curve is stable, while
  the dashed one is the `squeezed' solutions and
  unstable. All the quantities involved are in terms of the dimensionless
  variables defined in Eq.~\eqref{eq:dim-variables}. 
   {\hl Both curves are solutions of the scalar model with an attractive self-interaction. We fix $\tilde
   N = 5$ and $ \lambda=-1$. The upper (lower) curve has a radius of
  $\tilde R_{90}=1.2$ ($0.2$).}
}
  \label{fig:wf_shape_squeezed}
\end{figure}

\subsubsection{Stability of Boson Star Solutions}
\label{sec:analyt-appr-schr}
{\hl Boson stars result from different forces balancing against each other. When the self-interaction is negligible, a boson star solution exists due to a balance between kinetic pressure and gravity. If the self-interaction is non-negligible and is attractive, a star solution may arise from kinetic pressure balancing against self-interaction as well as gravity. If the dominant self-interaction is repulsive, the star solution could be due to repulsion plus kinetic pressure balancing against gravity. }

The stability of a solution can be numerically verified by solving the
time evolution of the wave function. Stable solutions are those that
do not decay away at $t\rightarrow \infty$, when we add a small
perturbation at $t=t_0$, \textit{i.e.}
$\psi(t_0) \rightarrow \psi(\mathbf{x}, t_0) + \delta
\psi(\mathbf{x})$. Likewise, unstable solutions are those wave
functions that either collapse or blow up for very small
perturbations, $\delta \psi(\mathbf{x}) \ll \psi(\mathbf{x},
t_0)$. 
There is a difference in the asymptotic behaviors of the stable/unstable wave functions.
This is demonstrated in Fig.~\ref{fig:wf_shape_squeezed}. From the figure, one can see that at large $r$,
the stable wave function falls off as $\mathrm{e}^{-r/R}$ while the unstable wave function is {\hl more squeezed with the asymptotic behavior $\sim 1/r$}. 

To obtain an analytical understanding of the solution stability, we take the one variable parametrized exponential ansatz:\footnote{This ansatz does not satisfy the regularity condition $\psi^\prime(0) = 0$. Yet it will not affect our discussion. One can modify this ansatz a bit to preserve the regularity condition, e.g., change it to the linear-exponential ansatz $\psi(r)\sim (1+r/R) \mathrm{e}^{-r/R}$. For comparison of different ansatzes, c.f. \cite{Eby:2015hsq, Schiappacasse:2017ham, Eby:2018dat}.}
\begin{eqnarray}
  \label{eq:exponential-ansatz}
  \psi(r) & =&
         \sqrt{\frac{N}{\pi R^3}} \mathrm{e}^{-r/R}, 
\end{eqnarray}
where $R \approx R_{90}/2.66$.
Plugging the ansatz into Eq.~\eqref{eq:NR-system-energy-components}, we find
\begin{eqnarray}
  H_{kin}
 & = &
   -   \frac{1}{2m}\int d^3 x \;  \psi(r) \nabla^2 \psi(r)
   = \frac{N}{2m R^2}   , \cr
   H_{int}
  & =& 
    \frac{    \lambda   }{4 f^2} \int d^3 x \; \psi(r)^4
    =      \frac{    \lambda     N^2}{32 \pi f^2 R^3}
, \cr
    H_{grav}
  & = &
    -Gm^2 \int_0^\infty \left (\int_0^{r}
    \psi(r) ^2
    \; 4\pi r^{\prime 2} \;dr' \right )
    \frac{1}{r}
    \psi(r) ^2
    \; 4\pi r^2 \;dr
    =
    - \frac{5 G_Nm^2N^2}{16 R},
    \label{eq:Hflat}
\end{eqnarray}
and the full Hamiltonian is 
\begin{align}
  H & =
      \frac{N}{2m R^2}       \mp \frac{|\lambda| N^2}{32 \pi f^2 R^3}          - \frac{5 G_Nm^2N^2}{16 R},
     \label{eq:HNR}
\end{align}
where the upper (lower) sign in the second term corresponds to the attractive (repulsive) case. 

\begin{figure}[t]
  \centering
  \includegraphics[width=.45\textwidth]{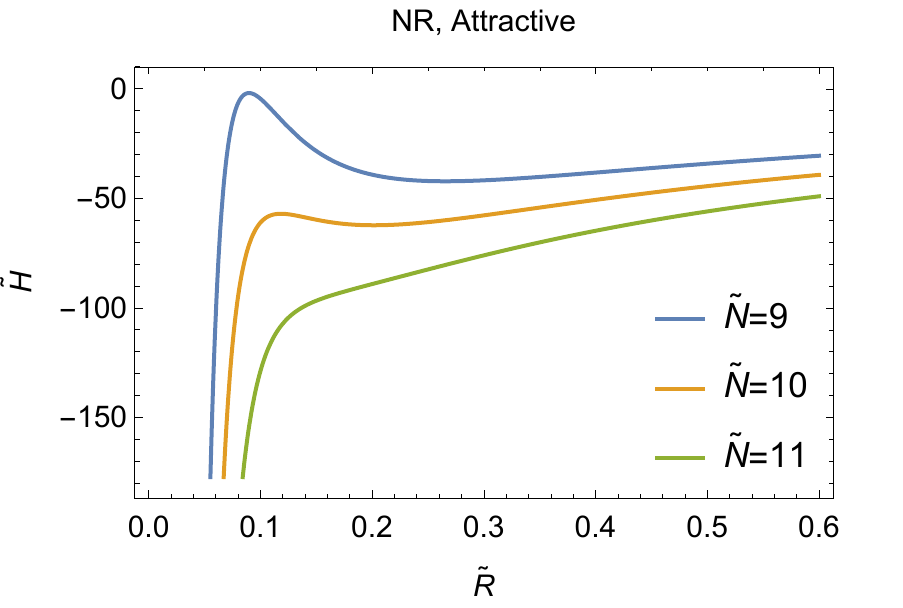
  }
  \includegraphics[width=.45\textwidth]{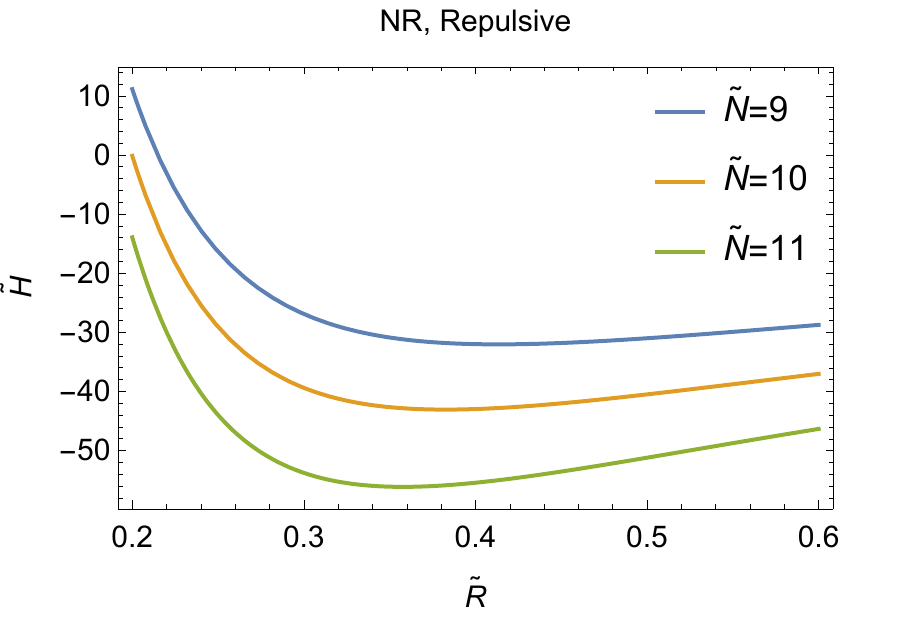
  }
  \caption{Left: for the attractive self-interaction, (rescaled) Hamiltonian $\tilde H$ as a function of the (rescaled) wave
    function width $\tilde R$. From $\tilde N=9$ to $\tilde
    N=10$, the local maximum moves to the right and the local minimum
    moves to the left. At $\tilde N=11$, there is no  local minimum,
    so there is no stable dark star with this $\tilde N$.
  Right: for the repulsive self-interaction, (rescaled) Hamiltonian as a function of the (rescaled)  wave
    function width $\tilde R$. There is always a minimum (corresponding to a stable solution) at any $\tilde{N}$. 
  }
  \label{fig:HR}
\end{figure}

For the attractive case (${\lambda}<0$), at a given $N$, when $R$ is large the gravity term dominates,
with $H(R) \sim 1/R \sim 0$; and
when $R$ is small the interaction term dominates, with $H(R)\sim -1/R^3
\sim -\infty$. In between the two limits, when there is a local minimum of the
energy at a finite $R$, there is a stable solution of the dark star. Given the two asymptotes, any local minimum (the stable solution) is accompanied
by a local maximum at a smaller $R$, which corresponds to an unstable solution. 
When $N$ increases, the local maximum moves to larger $R$ and the local minimum moves to smaller $R$. The limit in which the minimum and maximum are degenerate corresponds to the maximum value of $N$, beyond which no solutions are admitted.
This is illustrated in the left panel of Figure~\ref{fig:HR}. 

On the other hand, for the repulsive $|\phi|^4$ interaction, the sign of
the $1/R^3$ term is flipped. As a result, at large $R$, we still have
$H(R) \sim -1/R \sim 0$; however, at small $R$,
$H(R) \sim 1/R^3\sim +\infty$. This setup always admits a local
minimum but no local maximum, \textit{i.e.}, no unstable branch. This
is illustrated in the right panel of Figure~\ref{fig:HR}. More
detailed discussions on the stability of boson stars can be found in
\cite{Gleiser:1988rq,Jetzer:1988vr,Harko:2014vya,Eby:2016cnq,Levkov:2016rkk,Eby:2018zlv}.

\subsection{Gravitational Waves from Binary Merger Systems}
\label{sec:GWs}
In light of detecting mergers of boson stars, we will briefly outline the
basics of gravitational waves emitted in binary merger events and the LIGO sensitivity band.
During the inspiral phase of the merger, the GW frequency can be
expressed as
\begin{eqnarray}
  \label{eq:fGW}
  f_{GW}
  & = &
    \sqrt{ \frac{M_1 + M_2} {\pi^2 \ell ^3}}, 
\end{eqnarray}
where $\ell$ is the major semi-axis of the binary merger system, or
equivalently the separation of the two inspiral bodies (of masses $M_1$ and $M_2$) in the
circularly inspiral case. When the mass pair gets close to each other,
the point mass approximation breaks down and the inspiral phase ends. 
Assuming the two stars are of similar mass and size and following the
convention in Ref.~\cite{Giudice:2016zpa}, we define the innermost
stable circular orbit (ISCO) radius as a multiple of the star radius, 
\begin{eqnarray}
  \ell_{ISCO}
  & =&
    6 R, 
\end{eqnarray}
where we use $R$ to denote the star radius.
Plugging it back into Eq.~\eqref{eq:fGW}, we obtain the GW ISCO frequency
\begin{eqnarray}
  f_{ISCO}
  & =& 
    \sqrt{\frac{M_1 + M_2}{\pi^2 (6R)^3}}
    = \frac{C^{3/2}c^3}{2\pi \cdot 3^{3/2}G_NM},
\end{eqnarray}
where in the last equality we assume $M = M_1 = M_2$ for simplicity, and $C = G_NM/R$ denotes the compactness of the stars. As $f_{ISCO}$ can be taken as the peak frequency of the merger spectrum, and LIGO has the best signal-to-noise (SNR) ratio within the frequency band $50 \;\mathrm{Hz} \sim 1000\;\mathrm{Hz}$, one has the best hope of observing mergers with $50 \;\mathrm{Hz} \lesssim f_{ISCO} \lesssim 1000\;\mathrm{Hz}$, which gives the sensitivity band of LIGO in the $C-M$ plane. 

\begin{figure}[ht]
  \centering
  \includegraphics[width=.5\textwidth]{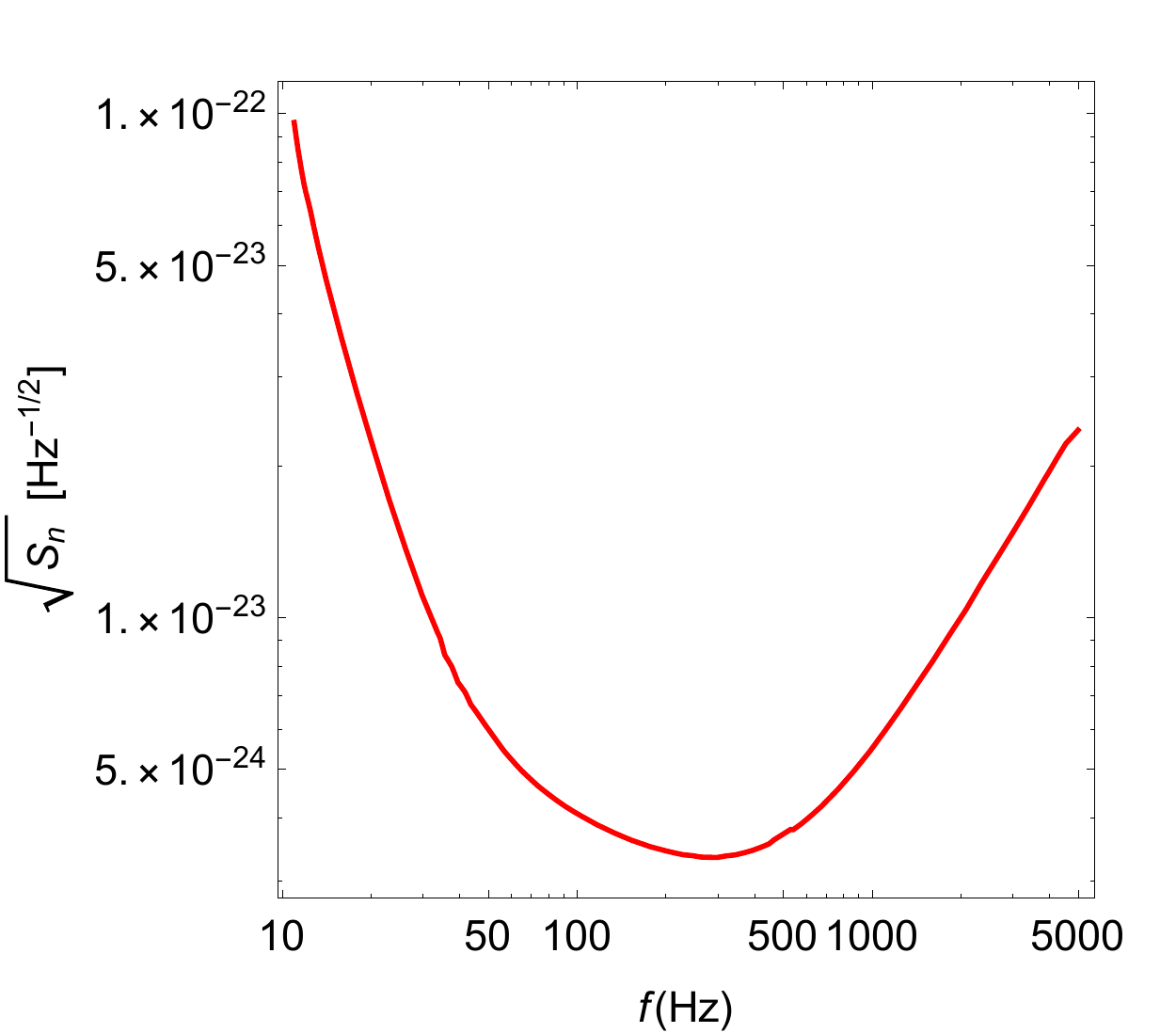
  }
  \includegraphics[width=.45\textwidth]{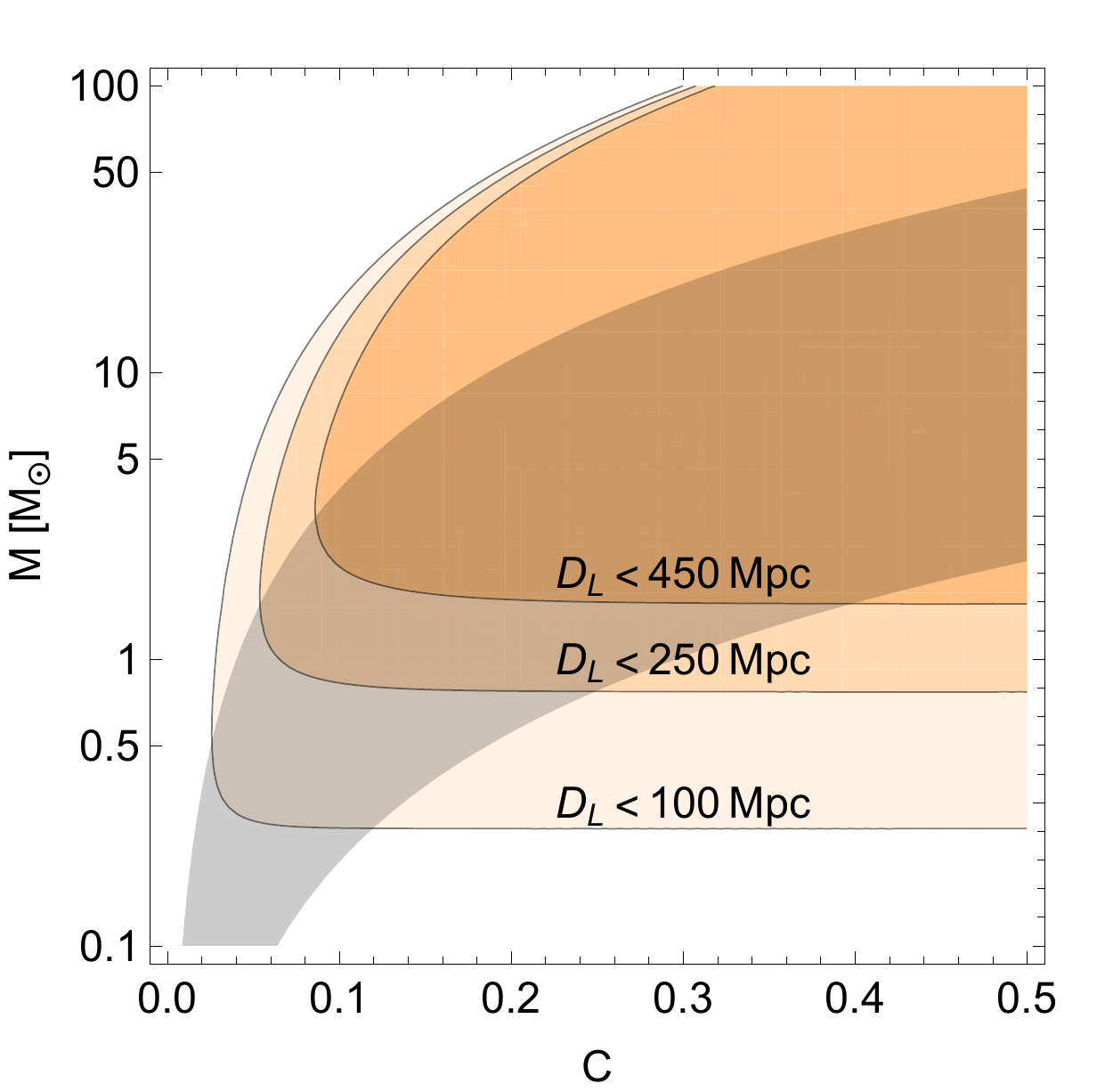
  }
  \caption{LIGO detector noise taken from Ref.~\cite{LIGONoise:2018}
    (left) and the sensitivity band in the $C-M$ plane (right). The gray band is obtained by demanding  $f_{ISCO}$ to be in the LIGO's best
  sensitivity range, 50 Hz $\sim$ 1000 Hz. The orange regions satisfy the detection criterion, $\rho>8$, for different ranges of luminosity distance $D_L$. 
}
  \label{fig:LIGOsensitivity}
\end{figure}

Besides the peak sensitivity determined by $f_{ISCO}$, one also needs
to consider the signal strength. 
Taking the leading order approximation of the quadrupole radiation,
the strain in frequency space is \cite{Khan:2015jqa}
\begin{eqnarray}
  \label{eq:strain-f}
  \tilde{h} (f)
  & \approx &
    \left (\frac{\sqrt{5/24}\; G_N^{5/6}}{\pi^{2/3} c^{3/2}} \right )
    \frac{M_c^{5/6}}{f_{GW}^{7/6} D_L}, 
\end{eqnarray}
where $D_L$ is the luminosity, and $M_c$ is the chirp mass, defined as
\begin{eqnarray}
  \label{eq:chirpmass}
  M_c & =&
        \frac{(M_1 M_2)^{3/5}}{(M_1 + M_2)^{1/5}}. 
\end{eqnarray}
Experimentally, the accumulation of the signal on top of the detector
noise is quantified by the SNR, which is defined as
\begin{eqnarray}
  \label{eq:SNR}
  \rho^2
  & \approx &
    4 \int ^{f_{ISCO}}_0 \frac{|\tilde{h}(f)|^2 }{S_n(f)} df,    
\end{eqnarray}
where $S_n(f)$ is the detector noise power spectral density \cite{LIGONoise:2018}. We require $\rho \geq 8$ for a possible detection of the signal \cite{Dominik:2014yma}. The
detector noise and the sensitivity bands in the $C-M$ plane are shown in Fig.~\ref{fig:LIGOsensitivity}. This band will be used later in the paper to identify parameters of pNGB models that can give rise to a detectable GW signal. 


\section{Effects of Nontrivial Curvature, Revisited}
\label{sec:re-visit-space}
In this section, we will solve the full equation of motion in GR for scalar dark matter with repulsive self-interaction, without making any approximations. We will see the GR corrections to the solution of the Schr\"odinger-Newton equation discussed in the previous section. It is known that scalar field theory in curved space-time is defined with certain ambiguities, such as the coefficients
of the $R \phi^2$ and $R_{\mu\nu\rho\sigma} R^{\mu\nu\rho\sigma} \phi^2$ terms, which vanish in the flat space limit \cite{Feynman:1996kb}. 
Nevertheless, we neglect such terms and follow the approach in
\cite{Colpi:1986ye}. We first review the main steps of the
computations in \cite{Colpi:1986ye} for the complex scalar field. 
The Lagrangian of a complex scalar with a repulsive $|\phi|^4$ interaction
reads 
\begin{equation}
  \label{eq:Lagrangian-KG-curved-space}
  \mathcal{L}=    \frac{1}{2}g^{\mu \nu}\nabla_\mu \phi^* \nabla_\nu \phi -
  \frac{1}{2}m^2 |\phi|^2 - \frac{\lambda}{4}\left
    (\frac{m^2}{f^2} \right )|\phi|^4,
\end{equation}
where $\nabla$ is the covariant derivative, and we use the $(+,-,-,-)$ signature. Note that in \cite{Colpi:1986ye}, the quartic interaction is simply written as $\lambda |\phi|^4$ {\hl while we use the parametrization in Eq.~\eqref{eq:model} motived by particle physics considerations. This will not affect any computations but make it clearer in the numerical computations for a given size of self-interaction, what the corresponding energy scale in the UV completion of the scalar model is.}
The energy momentum tensor is given by 
\begin{eqnarray}
  T_\mu^\nu
  & =&
        \frac{\delta \mathcal{L}}{\delta (\nabla_\nu \phi)}
    \nabla_\mu\phi
        +
    \frac{\delta \mathcal{L}}{\delta (\nabla_\nu \phi^*)} \nabla_\mu
    \phi^*
    -
     \delta^\nu_\mu \mathcal{L}    
    \cr
  & =&
       \frac{1}{2}     g^{\nu \nu'} \nabla_{\nu'} \phi^* \nabla_\mu
    \phi 
       + \frac{1}{2} g^{\nu\nu'} \nabla_{\nu'} \phi \nabla_\mu
    \phi^*
       -\delta_\mu^\nu \left ( \frac{1}{2} g^{\mu' \nu'}\nabla_{\mu'} \phi^*
       \nabla_{\nu'} \phi - \frac{1}{2}m^2 |\phi|^2 -
    \frac{\lambda}{4}\left
    (\frac{m^2}{f^2} \right )|\phi|^4 \right ).\cr
\end{eqnarray}
As before, we look for a boson star solution in a spherically symmetric metric: 
\begin{equation}
  \label{eq:metric}
  ds^2 =
    B(r) dt^2 - A(r) dr^2 - r^2 d\theta^2 - r^2 \sin^2 \theta d
    \phi^2. 
\end{equation}
The Einstein tensor $G_\mu^\nu$ is diagonal, with the following non-zero
components:
\begin{eqnarray}
G_t^t& =&   -\frac{A'(r)}{r A(r)^2}+\frac{1}{r^2
       A(r)}-\frac{1}{r^2},
       \cr
       G_r^r
  & =& \frac{B'(r)}{r A(r)
   B(r)}+\frac{1}{r^2
               A(r)}-\frac{1}{r^2},
    \cr
    G_\theta^\theta
    & = &
               -\frac{A'(r) B'(r)}{4
   A(r)^2 B(r)}-\frac{A'(r)}{2 r
   A(r)^2}+\frac{B''(r)}{2 A(r)
   B(r)}-\frac{B'(r)^2}{4 A(r)
   B(r)^2}+\frac{B'(r)}{2 r A(r)
      B(r)},
      \cr
      G_\phi^\phi
  & =&
    -\frac{A'(r) B'(r)}{4 A(r)^2
   B(r)}-\frac{A'(r)}{2 r
   A(r)^2}+\frac{B''(r)}{2 A(r)
   B(r)}-\frac{B'(r)^2}{4 A(r)
   B(r)^2}+\frac{B'(r)}{2 r A(r) B(r)}.
\end{eqnarray}
Solving the ${}^t_t$ and ${}^r_r$ components of Einstein equation, we
have
\begin{align}
  \label{eq:KG-eom-1-2}
& \frac{4\pi G_N}{B(r)} \partial_t \phi \partial_t \phi^*
  +  \frac{4\pi G_N}{A(r)} \partial_r \phi \partial_r \phi^*
                          +4\pi G_N m^2 |\phi|^2 + 2G_N \pi \lambda
                          \left (\frac{ m^2}{ f^2}\right ) |\phi|^4 - \frac{A'(r)}{r
  A(r)^2} + \frac{1}{r^2 A(r)} - \frac{1}{r^2} = 0,
  \cr
& \frac{4\pi G_N}{B(r)} \partial_t \phi \partial_t \phi^*
  +  \frac{4\pi G_N}{A(r)} \partial_r \phi \partial_r \phi^*
        -4\pi G_N m^2 |\phi|^2 - 2G_N \pi \lambda
                          \left (\frac{ m^2}{ f^2}\right )
        |\phi|^4
-\frac{B'(r)}{r A(r) B(r)} - \frac{1}{r^2 A(r)} + \frac{1}{r^2}
        = 0.
\end{align}
It is noted
that the twice contracted Bianchi identity, $\nabla_\mu G^\mu_\nu =
0$, is satisfied automatically. 
One extra constraint comes from the scalar EOM, $\nabla_\mu
\nabla^\mu \phi - m^2 \phi - \frac{\lambda m^2}{f^2} |\phi|^2 \phi = 0$. 
Plugging in the covariant derivative, we find
\begin{align}
  \label{eq:KG-eom-3}
  \frac{1}{A}\partial_r^2 \phi - \frac{1}{B}  \partial_t^2 \phi +
  \partial_r \phi \left (\frac{B'(r)}{2 A(r)B(r)} - \frac{A'(r)}{2A(r)^2} +
  \frac{2}{A(r)r}\right ) - m^2 \phi - \lambda\left (
  \frac{m^2}{f^2} \right ) |\phi|^2 \phi = 0.
\end{align}
Eqs.~\eqref{eq:KG-eom-1-2} and \eqref{eq:KG-eom-3} together form the
Einstein-Klein-Gordon system. To solve the system, we adopt the harmonic ansatz 
\be \phi(r, t) = \Phi(r) \mathrm{e}^{-i \mu t} 
\label{eq:haran}
\ee 
and make the
following rescalings:
\bea
  r &  =& \tilde r \; \left ( \frac{1}{m} \right ), \quad
      \Phi  = \tilde \Phi \; (4\pi \; G_N)^{-1/2} , \cr
             \mu & =& \tilde \mu \; m, \quad \quad \quad 
                   \lambda  = \tilde \lambda \; 
                             (4\pi \; G_Nf^2),
\eea
where the variables with tildes are dimensionless as before.
Then the Einstein-Klein-Gordon system becomes
\begin{align}
  \label{eq:KG-eom-rescaled}
  &
    \left ( \frac{\tilde \mu^2}{B} + 1 \right )  \tilde
    \Phi^2 
    +  \frac{1}{A}  {\tilde \Phi^{\prime 2}}
    + \frac{1}{2}\tilde \lambda \tilde
    \Phi^4    - \frac{A'}{\tilde r  A^2} +
    \frac{1}{\tilde r^2 A} -
    \frac{1}{\tilde r^2} = 0, 
  \cr
& \left ( \frac{\tilde \mu ^2 }{B}  - 1 \right )\tilde \Phi^2
  +  \frac{1}{A} \tilde  \Phi'^2
        - \frac{1}{2} \tilde \lambda \tilde \Phi^4
-\frac{B'}{\tilde r A B} -
        \frac{1}{\tilde r^2 A} + \frac{1}{\tilde r^2}
        = 0,
        \cr
  &
    \frac{1}{A}{\tilde  \Phi^{\prime\prime }}
    +\left (\frac{\tilde \mu^2}{B} - 1 \right )\tilde \Phi +
  \tilde \Phi' \left (\frac{B'}{2 AB} - \frac{A'}{2A^2} +
  \frac{2}{A \tilde r}\right )  - \tilde \lambda \tilde \Phi^3 = 0,
\end{align}
where the prime indicates a derivative with respect to $\tilde r$.

\subsection{ADM Mass and Local Energy Density 
}
\label{sec:solutions-psi-boson}
We solve the equations using the shooting method with
\texttt{Mathematica}, and verify the solutions by collocation
method\footnote{Please note that when using collocation method to
  solve this system, one encounters a so called \textit{singular
    boundary value
    problem} \cite{de1976difference,weinmuller1984boundary,weinmuller1986collocation,burkotova2016asymptotic,burkotova2018nonlinear}. In
  order to solve such problems, the boundary condition needs to be
  carefully prepared. Mathematically, it is proven that the shooting
  is not well suited for this kind of problem as it does not guarantee a
  unique solution \cite{koch2003convergence,SBVPshooting}.} using \texttt{BVPSUITE}
\cite{kitzhofer2009bvpsuite,kitzhofer2010new}.  After obtaining the
solution of $\Phi$, one can easily compute the boson star profiles,
such as mass, radius and compactness. The boson star mass is given by
the ADM mass, which is defined asymptotically at spatial infinity \be
  \label{eq:ADMmass}
 M = \frac{1}{2G_N}
        \lim_{r\rightarrow \infty} \left ( 1- \frac{1}{A(r)} \right ) r.
\ee
Using Eq.~(\ref{eq:KG-eom-rescaled}), one can show that this is equivalent to 
\be
    M
  =\int_0^\infty dr  \;  4\pi r^2\;
    T_0^0, 
\ee
with the $00$-th component of the energy-momentum tensor given by 
\bea
  T_0^0
  & = &
   \frac{\delta \mathcal{L}}{ \delta \partial_0 \phi} \partial_0 \phi
    - \mathcal{L} 
    \cr
  & = &
    \frac{\mu^2}{2B}\Phi^2 
    +\frac{1}{2}m^2 \Phi^2
    +\frac{1}{2A}(\partial_r \Phi)^2
    + \frac{\lambda}{4} \left ( \frac{m^2}{f^2}\right ) \Phi^4 
        \label{eq:energydensity}
\eea
{\hl A general review of ADM formalism can be found in~\cite{Arnowitt:1962hi}.}

\begin{figure}[h]
  \centering
  \includegraphics[width=.5 \textwidth]{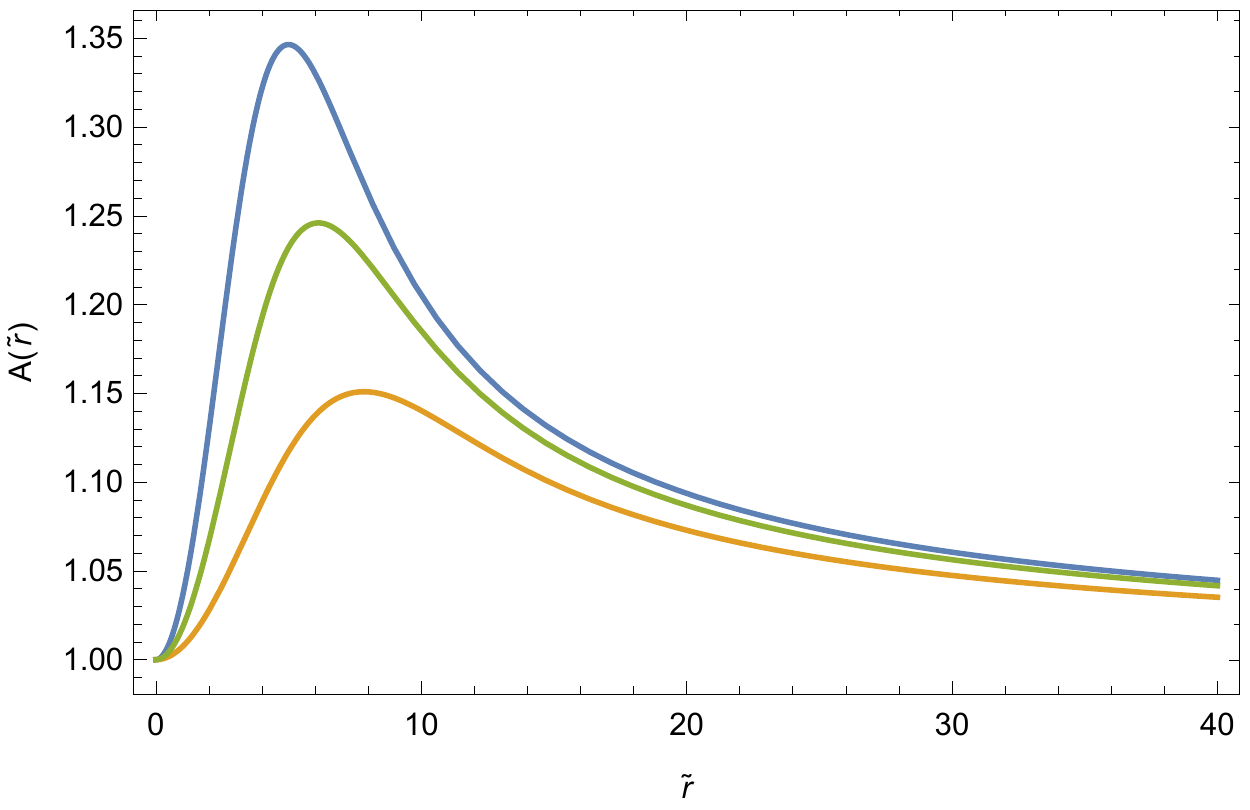
  }
  \caption{A few benchmarks solutions of the $g_{rr}$ component, $A( \tilde{r})$, as a function of the
    dimensionless variable $\tilde{r} $. We fix $f=5\times
    10^{17}\; \mathrm{GeV}$, with $\tilde{\phi}(0)=0.2,0.15,0.1$, from
    top to bottom.}
  \label{fig:A}
\end{figure}

Note that the ADM mass can not be interpreted simply as a volume integration of energy density. In curved space, the proper volume integration should include the induced spatial metric: 
\be
\int_0^\infty dV  \sqrt{-g_{\rm ind}}  \; T_0^0 =    \int_0^\infty dr  \;\sqrt{A(r)}\;  4\pi r^2\; T_0^0
\ee
In the weak gravity limit, \textit{i.e.}, $A(r) \approx B(r) \approx 1$, the volume integration above is approximately the ADM mass. Yet numerically, we observe that $A(r)$ may deviate significantly from $1$, as shown in Fig.~\ref{fig:A}. The two integrations may differ by $\sim 15$\% or more. An explanation is that the difference between the two integrations is the gravitational binding energy inside the star~\cite{Carroll:2004st}.

\subsection{Curvature Effect on Boson Star Mass Profile}
\label{sec:comp-einst-klein}
\begin{figure}[h]
  \centering
  \includegraphics[width=.55\textwidth]{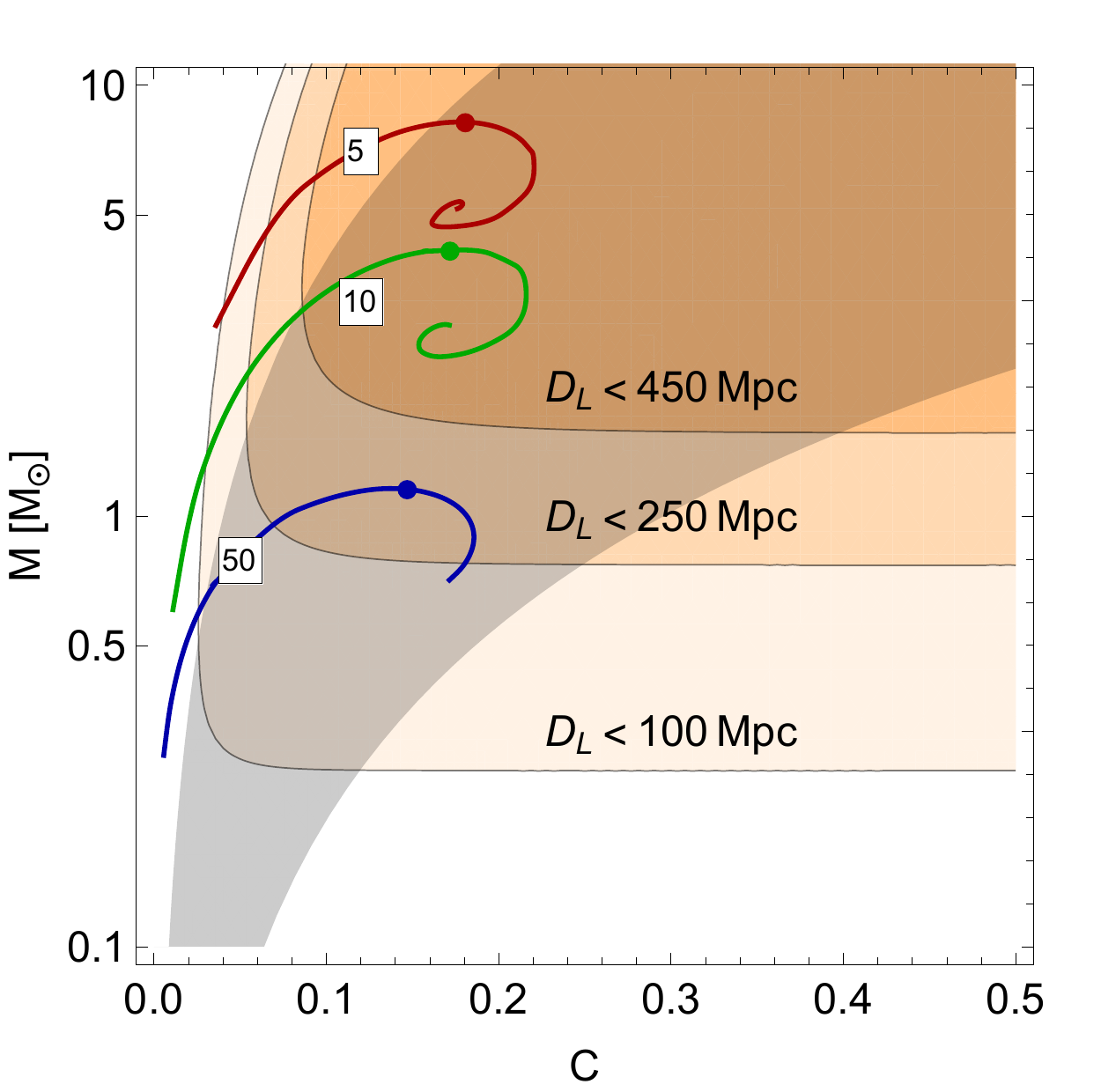}
  \caption{Contours of boson star solutions from the $|\phi|^4$ theory
    in the $C-M$ plane. The shaded regions are the same as presented
    in Fig.~\ref{fig:LIGOsensitivity} of Section~\ref{sec:GWs}.  The
    gray shaded region corresponds to $f_{ISCO}$ in the LIGO
    sensitivity band, and the orange regions have SNR above the
    detection threshold for different ranges of luminosity
    distance. All of the three benchmark curves are generated for
    $m= 10^{-10} \; \mathrm{eV}$ and $\lambda =
    1$. 
    Each curve is labeled by $f/(10^{16}\;\mathrm{GeV})$. The turning
    points that correspond to the maximal mass are marked with a
    dot. }
  \label{fig:CMPlot}
\end{figure}

We fix $\lambda$ and $m$ and vary $f$ to change the strength of the quartic coupling. For different $f$, we compute the compactness and mass of the solutions to Eq.~(\ref{eq:KG-eom-rescaled}), which is presented in Fig.~\ref{fig:CMPlot}. From it, one can see that for each repulsive quartic coupling, there
are two branches of solutions with a turning point at the maximum mass with roughly a {\it common} compactness:
  \bea
  M_{\rm max}  &\approx &  3 M_\odot \, \sqrt{\lambda} \, \left(\frac{10^{17} \, {\rm GeV}}{f}\right) \left(\frac{10^{-10} \,{\rm eV}}{m}\right),\\
  C_{\rm max} &\approx & 0.16.
  \label{eq:MC}
  \eea
Based on dimensional analysis, $M_{\rm max} \sim M_{\rm pl}^2/m$. $M_{\rm max}$ is also proportional to the square root of the scalar's self-coupling. Since we parametrize the self-coupling as $\lambda m^2/f^2$ (motivated by identifying the scalar as a pNGB), $M_{\rm max} \propto 1/f$. The branch to the left of the turning point is stable while the right branch is unstable. 

\begin{figure}[h]
  \centering
  \includegraphics[width=.5\textwidth]{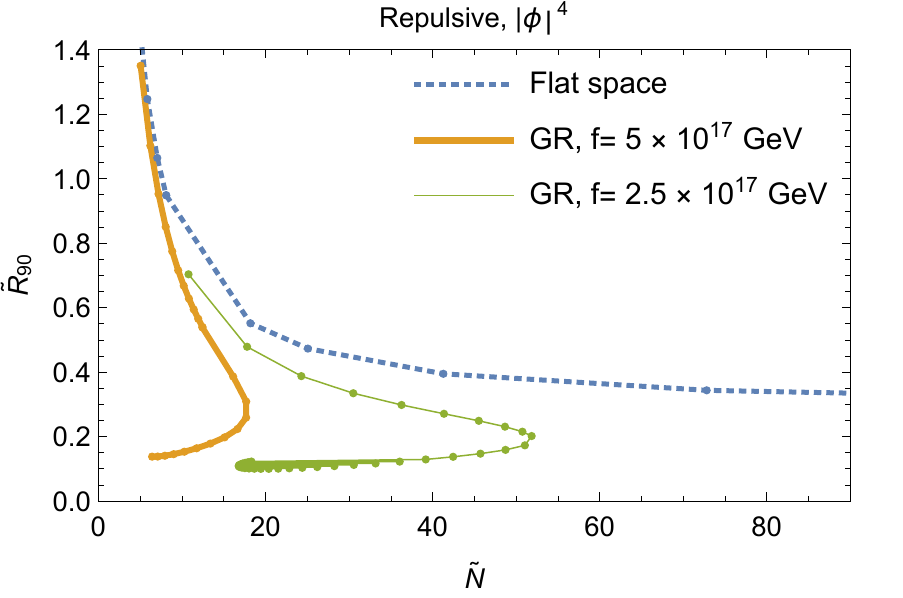
  }
  \caption{Comparison between the NR flat space solution
    (dotted blue) and the GR solution for the repulsive quartic interaction,
    $f = 5\times 10^{17}\; \mathrm{GeV}$ (thick orange), and $2.5\times
    10^{17}\; \mathrm{GeV}$ (thin green). We fix the scalar mass and
    $\lambda$ to be the same. The dimensionless variables are defined
    in Eq.~\eqref{eq:dim-variables}. They are chosen
    such that they are  scale independent in the NR flat space solution. 
    \label{fig:NR90-GR-NR}}
\end{figure}

To demonstrate the difference between the GR solution and the NR flat space solution presented in Fig.~\ref{fig:SN-attr-vs-repul}, we also present the numerical results in the rescaled particle number and $R_{90}$ plane in Fig.~\ref{fig:NR90-GR-NR}. Again we observe that in GR, there are two branches of solutions for fixed repulsive quartic interaction and there exists a maximal particle number beyond which there is no stable star solution. This is clearly different from the solution in the NR flat space limit, which extend to arbitrarily large particle number. 

To see where the turning point comes from, we will work in the weak gravity limit but keep the leading order perturbation, $V(r)$, to the flat-space metric
\be
  \label{eq:metric-perturb}
    ds^2 =
    -(1+2V(r)) dt^2 + (1-2V(r)) dr^2 + r^2 d\theta^2 + r^2 \sin^2 \theta d
    \phi^2,
\ee
We adopt the following set of ansatz, 
\bea
  \label{eq:phi-V-ansatz}
\Phi(r)&=&
         \sqrt{ \frac{N}{\pi m R^3}} \; \mathrm{e}^{-r/R}, \cr
         V(r) & =& - \frac{G_NM(r)}{r}.
\eea
$M(r)$ is defined as follows:
\be
  \label{eq:sc-ansatz}
  M(r)
   \equiv
    \int_0^r   
    \; T_0^0\; 4\pi r'^2 dr'
    = 
    \int_0^r  
    m^2 \Phi^2(r) \; 4\pi r'^2 dr', 
\ee
which is the mass enclosed within the sphere of
radius $r$. Since $T_0^0 \approx m^2 \Phi^2$ up to corrections of higher order in $V$ and $M(r)$ appears in $V$, we drop higher order terms in defining $M$.  
Plugging Eq.~\eqref{eq:phi-V-ansatz} into $M(r)$, we have 
\bea
  V(r) & =&
      -      \frac{G_Nm^2}{r}
      \int_0^r \Phi(r')^2 4 \pi r^{\prime 2} \; dr'
    \cr
  &  =&
    -      \frac{G_NmN}{r}
    \left (
    1- \mathrm{e}^{-2r/R} \left( 1+ \frac{2r}{R} + \frac{2r^2}{R^2}\right) \right ). 
\eea
{
  The set of ansatz matches the asymptotes of the numerical solutions but deviates at small $r$.}

Applying the ansatz into the energy density in Eq.~\eqref{eq:energydensity} and carrying out the volume integration, we find that the first two terms in $T^0_0$ contribute
\bea
H_{mass}+  H_{grav} 
  & =&
    \int_0^\infty \left ( \frac{\mu^2}{2B}\Phi^2 + \frac{m^2}{2}\Phi^2 \right )
    \; 4\pi r^2 \; dr
    \cr
  & =&
    \int_0^\infty  \left (   (1-2V) 
\frac{\mu^2}{2}\Phi^2 +
\frac{m^2}{2} \Phi^2 \right )  \; 4\pi r^2
    \; dr
    \cr
  & =&
    mN - \frac{5 G_N m^2 N^2}{16 R}. 
  \label{eq:Hgravcurved}
\eea
The first term is simply the rest mass of $N$ scalars with mass $m$ while the second term is the gravitational potential energy, matching the result in the flat space limit in Eq.~\eqref{eq:Hflat}. {
  In the derivation above, we consider the leading order gravitational correction to $\mu^2$ and parametrize it as 
\be
\mu^2 = m^2 \left(1 - \alpha \frac{G_N m N}{R}\right), 
\ee
where $\alpha= 5/4$ to reproduce the gravitational potential energy in Eq.~\eqref{eq:Hflat}. Numerically we observe that the deviation of $\mu^2/m^2$ from 1 is anti-correlated with $G_N m N/R \propto C$ and is roughly the same. Higher order corrections are of order ${\cal O}(G_N^2 m^2 N^2/R^2)$. }

The kinetic energy in the curved space, $(\partial_r \Phi)^2/(2A)$ contributes
\bea
H_{kin} 
   =
    \int_0^\infty \frac{1}{2A} \left (\partial_r \Phi \right)^2  
    \;  4 \pi r^2 dr
    =
    \frac{N}{2m R^2}
    - \frac{5 G_NN^2}{16 R^3}.
\eea
The first term is the same as the kinetic energy term in the flat space in Eq.~\eqref{eq:Hflat}. The extra term is due to non-trivial spatial curvature and is $\propto -1/R^3$ and is absent in Eq.~\eqref{eq:Hflat}. 

Lastly, the interaction energy with the ansatz gives
\begin{equation}
  H_{int}=
    \int_0^\infty \frac{\lambda}{4} \left ( \frac{m^2}{f^2} \right ) \Phi(r)^4 
4\pi r^2 \; d r = 
\frac{\lambda  N^2 }{32 \pi f^2 R^3}. 
\end{equation}
Note that the $H_{int}$ above is for complex scalars. For real scalars, with the fast oscillation modes neglected, the interaction term can be obtained by replacing $\Phi^4$ with $3\Phi^4/2$,
which matches exactly with the NR result in Eq~\eqref{eq:Hflat}.

Putting everything together, in the repulsive $\Phi^4$ theory, the Hamiltonian of the dark star reads,
\bea
  M(R)
  & =& H_{mass} + H_{kin} + H_{grav} + H_{int} \nonumber \\
  &= & mN +
 \frac{N}{2m R^2} - \frac{5G_N m^2N^2}{16R} + \left (
        \frac{\lambda }{32 \pi f^2} {
          - \frac{5G_N}{16}} \right ) \frac{N^2}{R^3}  
      \eea
Compared to the result in the flat space limit in Eq.~\eqref{eq:HNR}, the additional term $\propto -5G_N/16$ can flip the sign of $1/R^3$ when the repulsive self-interaction is not large. If the sign of the $1/R^3$ term is negative, stable solutions are always accompanied with unstable ones, following the same argument for attractive self-interaction in Sec.~\ref{sec:analyt-appr-schr}. There is an allowed maximum particle number, hence boson star mass, when these two branches meet. This feature originates from the nontrivial curvature of the background space, which the Schr\"odinger-Newton equation fails to capture. 

We can also do a rough estimate of the maximal compactness for the repulsive boson star. When $1/R^3$ term is negative, solving $\partial H /\partial R=0$, we find two solutions, one stable at $R=R_1$, and one unstable at $R=R_2$, with $R_2<R_1$. Requiring the two solutions to be degenerate, we get the maximum boson star mass. This happens at 
\be
R_1=R_2 \equiv R_0  = \frac{8}{5 G_N m^3 N}. 
\ee
The corresponding particle number is 
\begin{align}
  N_{max}^2 & =
        \frac{256 \pi f^2}{300 \pi G_N^2 m^4 f^2 - 30 G_N m^4\lambda}.
\end{align}
This gives the maximal compactness: 
\begin{align}
  C_{max} = \frac{G_N M_{90}}{R_{90}} = 
  \frac{G_N m (0.9 N_{max})}{ 2.66 R_0}
    \approx 0.18,
\end{align}
in the limit of large $f$. We have used the fact that $R_{90}$ is $2.66 R_0$ given the exponential ansatz of the wave function. This is consistent with the numerical result in Eq.~\eqref{eq:MC}.

The analytic derivation based on ansatz here should only be taken as a heuristic argument to shed some light on the origin of the maximum compactness. 
Even in this crude argument, when $10 \pi G_N f^2 < \lambda$ or equivalently $f \sim 10^{17}$ GeV when $\lambda \sim {\cal O}(1)$, the curvature correction cannot be as important as the self-interaction and flip the sign of $1/R^3$ term. {
  In addition, our whole argument is based on another ansatz, the
  single harmonic ansatz in Eq.~\eqref{eq:haran}.} One might wonder
whether for larger repulsion with smaller $f$, the turning point
disappears. Our numerical solver no longer works when
$\tilde{\lambda}$ is as large as $\mathcal{O}(10^4)$ since the system
becomes highly non-linear. Thus it remains an open question what
happens when the repulsive self-interaction is large and whether there
are stable dark stable solution with larger compactness. We leave this question
for future work. For a different treatment using hydrodynamic
approach, see \cite{Chavanis:2011cz}.

\section{Realistic Light Scalar DM Model}
\label{sec:realistic-dark-star}
For bosons to behave collectively and form a BEC, the occupancy number of the scalars has to be much larger than 1. 
This is equivalent to requiring the de Broglie wavelength $2\pi/(mv)$ to be greater than the inter-particle separation $(m/\rho)^{1/3}$. Taking the average dark matter density in the Milky Way, this gives an upper bound on the scalar mass $m \lesssim 1 \;\mathrm{eV}$.\footnote{This upper bound is crude and could be relaxed if the initial density is much larger when the light scalars start to clump.} To avoid large scale structure constraints, the scalar cannot be too light either. Its de Broglie wavelength has to be at most comparable to the size of a dwarf galaxy $2\pi/(mv)\lesssim 1 \; \mathrm{kpc}$, leading to a lower bound on the scalar mass $m \gtrsim 5 \times 10^{-23} \; \mathrm{eV}$. Thus we only consider very light scalars in the mass range from $10^{-23}\; \mathrm{eV}$ to $1 \; \mathrm{eV}$.

In a more realistic particle physics model, the radiative stability of the mass needs to be addressed for such light scalars. One way is to identify the scalar as a weakly-coupled pNGB and protect its mass by an approximate shift symmetry.\footnote{Another possibility is that the light scalar is a composite object such as a glueball inside a low-scale confining hidden sector~\cite{Soni:2016gzf}.} 
The scalar mass is an order parameter for the explicit breaking of the associated UV symmetry and the potential is more complicated than the $|\phi|^4$ model. In the following, we illustrate this with a concrete model and show the corresponding boson star solutions.

\subsection{A pNGB Model with a Repulsive $|\phi|^4$ Interaction}

In general, it is demonstrated in~\cite{Fan:2016rda} that it is non-trivial to construct a pNGB model with repulsive self-interaction in the NR limit. For example, for the classic QCD axion model, taking the phenomenological potential and expanding it around the minimum, 
\bea
    V(\phi)  &=&   \Lambda^4
    \left ( 1- \cos \left( \frac{\phi}{f} \right)\right ) = 
    \frac{\Lambda^4}{2f^2} \phi^2 - \frac{1}{4!} \frac{\Lambda^4}{f^4}
    \phi^4 + ...
    \cr
  & =&
    \frac{m^2}{2} \phi^2 - \frac{1}{4!} \frac{m^2}{f^2} \phi^4 + ...
\eea
one finds that the leading order quartic interaction is attractive. 
In general, for an axion-like particle from breaking of a compact symmetry and with a single trigonometric potential, the leading interaction term is always attractive. Yet there is no no-go theorem. 
One concrete example of a pNGB with leading repulsive self-interaction is a $5D$ gauged $U(1)$ theory with the fifth dimension compactified on a circle~\cite{Fan:2016rda}. The pNGB is the gauge invariant Wilson loop of the fifth component of the gauge field. It obtains a one-loop effective potential from 5$D$ charged matter. {\hl This example serves as a concrete UV completed particle physics model for us to compute the boson star solution. The complication is that in this model, the pNGB is real. Ignoring the fast oscillation modes, the procedure to find a boson star solution is very similar mathematically for both real and complex scalars. Yet the relativistic star from real scalars could be cosmologically unstable due to particle number changing process such as the $3 \to 1$ process~\cite{Eby:2015hyx}.\footnote{Estimate of the boson star lifetime taking into account of gravity and the non-linear effects requires simulations analogous to that for QCD axion~\cite{Helfer:2016ljl}, which is beyond the scope of the paper.} Thus the relativistic stars may not have any observational consequences. Nevertheless, one can still find boson star solutions by solving equation of motion and we will present the results in Appendix~\ref{sec:cosine-potential-real}. One can see that the boson star solutions can be very different when solving a full cosine potential, compared to only keeping the leading interaction.}

 {\hl Below, to avoid the lifetime complication, we take a phenomenological complex scalar potential, different from the $|\phi|^4$ model 
\begin{align}
  \label{eq:pheno-potential}
  V(|\phi|)
  & =
    \frac{1}{2}m^2 |\phi|^2 + \frac{1}{2Q^2}m^2 f^2 \left [1-\cos \left (\frac{Q |\phi|^2}{f^2} \right ) \right ], 
\end{align}
where $Q$ is a dimensionless variable that parametrizes a family of the cosine potentials, which all have the same $|\phi|^4$ interaction when one expands the potential around $\phi = 0$. This model is again motivated by identifying the scalar as a pNGB. If the associated symmetry is compact, the pNGB's potential can be a single cosine function (or a combination of oscillatory terms), with the dimensionless quantities in the potential determined by the parameters in the underlying particle physics models. We do not intend to construct a full model for a complex pNGB here, which would be an exercise similar to the composite Higgs scenarios.
 }

\subsection{Numerical Star Solution of the Cosine Potential}
\label{sec:numer-solut-cosine}

\begin{figure}[h]
  \centering
  \includegraphics[width=.45\textwidth]{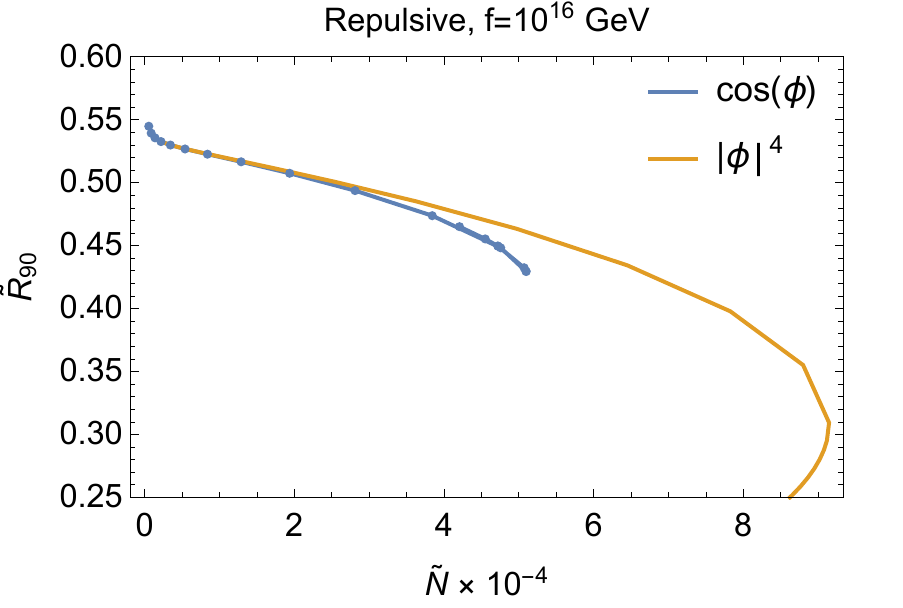}
  \caption{Comparison of the radius and particle number between
    $|\phi|^4$ potential and the cosine potential in Eq.~\eqref{eq:pheno-potential}. 
    We have chosen $f = 10^{16}\; \mathrm{GeV}$ and $Q=5$ as a benchmark. The coefficient of the $|\phi|^4$ interaction is chosen to match the leading
    term in the cosine potential. {\hl This dimensionless variable
      $\tilde R$ and $\tilde N$
      scales with $1/m$ and $1/m^2$ respectively.}  }
  \label{fig:phi4-cosine}
\end{figure}

\begin{figure}[h]
  \centering \includegraphics[width=.55\textwidth]{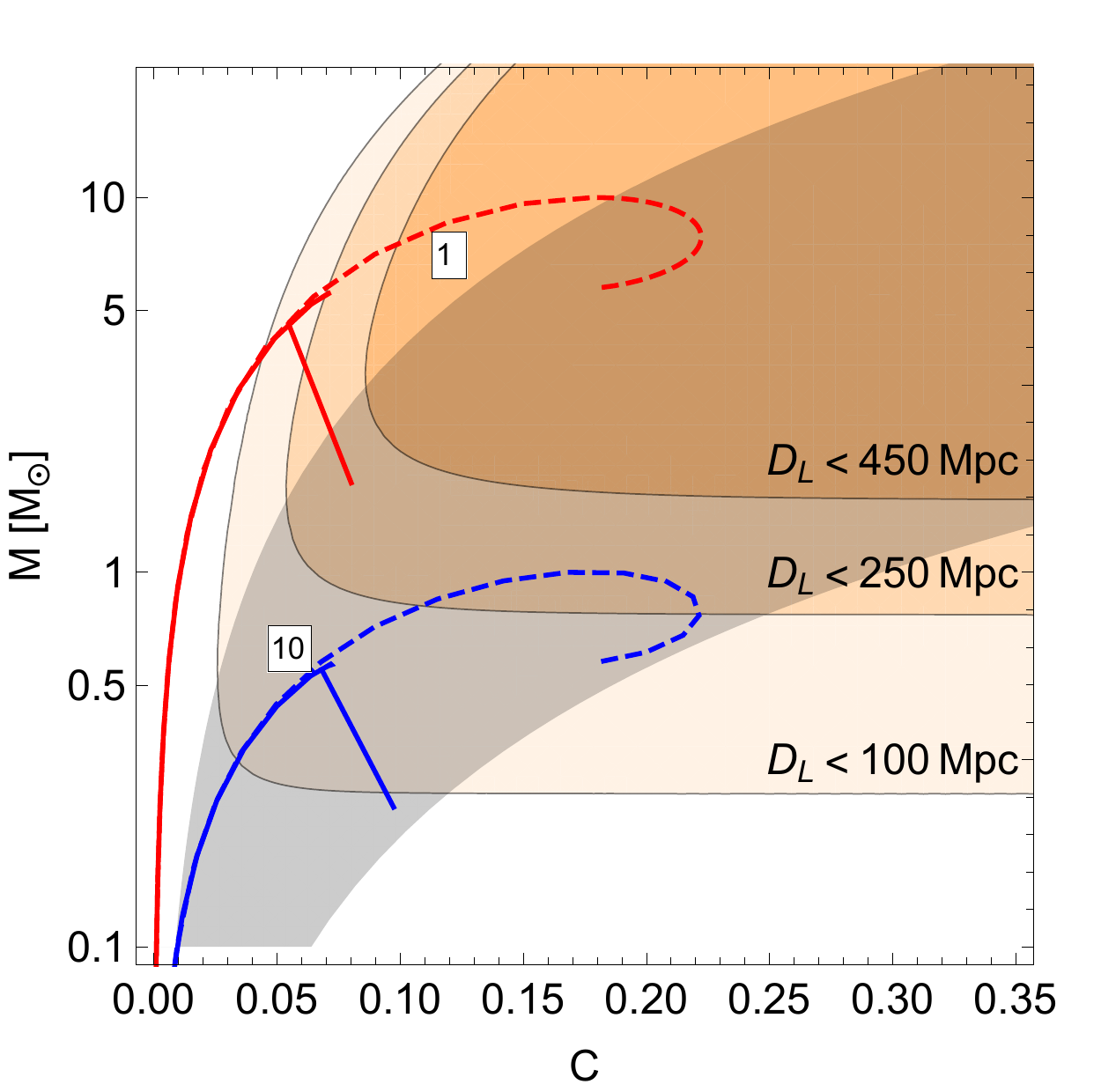
  }
  \caption{The compactness and mass profiles of boson stars for the
    cosine potential model (solid) and the $|\phi|^4$ model (dashed)
    for $f=10^{16}\;\mathrm{GeV}$, $m= 10^{-9} \;\mathrm{eV}$.  The
    cosine potential is shown in Eq.~\eqref{eq:pheno-potential} with
    $Q=5$ as a benchmark point. The $|\phi|^4$ interaction is chosen
    to match the leading term in the expansion of the cosine
    potential. The end point of the curve happens when the phase of
    the potential is of order one. The shape is due to the stiffness
    of the numerical system during the transition and finite
    resolution between neighbor points. After crossing the stiff
    region, the solution is verified to be numerically stable. The
    shaded region corresponds to $f_{ISCO}$ falling into the LIGO
    sensitivity band, and the orange regions are the luminosity
    distance of such mergers in order to have SNR above the detection
    threshold.  See Section~\ref{sec:GWs} for details. }
  \label{fig:CM-phi4-cos}
\end{figure}

The boson star solutions of the complex scalar theory {\hl with the cosine potential in Eq.~\eqref{eq:pheno-potential}} in the rescaled particle number and star radius plane are presented in Fig.~\ref{fig:phi4-cosine}. In the same figure, we also present the star solutions from the simple $|\phi|^4$ potential. We choose the two potentials to agree at the order $|\phi|^4$ in the expansion of the cosine.
It is observed that when $f$ decreases, higher-order interaction terms become more important and can lead to different phase diagrams for these two different potentials. {\hl One technical note is that when higher-order terms are important, the equation of motion becomes more non-linear and it becomes increasingly difficult to find a solution. In our study, we push our differential equation solver to the boundary of the parameter space in which it works reliably.} We also present the solutions in the star mass and compactness plane in Fig.~\ref{fig:CM-phi4-cos} and show the results from $|\phi|^4$ theory for comparison as well. 

{\hl In producing Fig.~\ref{fig:phi4-cosine}, it is observed that the
solutions to the cosine potential reaches a maximal particle number
when the nonlinear effect becomes significant, i.e., the argument in the $\cos$ potential becomes an order one number.  After that, the
particle number decreases. This is quite different from the
predictions of the $|\phi|^4$ theory. Note that the difference may become bigger
when $f$ is lower and the scalar self-interaction becomes stronger, or
when the charge $Q$ is larger, which does not affect the leading
$|\phi|^4$ expansion.}
For a fixed $f$ (equivalently fixed quartic interaction), the maximum mass and compactness shrinks in the pNGB model with the cosine potential as shown in Fig.~\ref{fig:CM-phi4-cos}. The reason for the decreasing compactness is that the sub-leading $|\phi|^8$ interaction in the cosine potential in Eq.~\eqref{eq:pheno-potential} is attractive. In particular, there is no longer a common maximum compactness for cosine potentials with different sizes of self-interactions (i.e., different $f$'s), in contrast to the $|\phi|^4$ potential. 

This also restricts the allowed range of scalar self-interaction that can give rise to sizable boson stars, whose binary mergers can be detected. As shown in Fig.~\ref{fig:CM-phi4-cos}, for a $10^{-9}$ eV
pNGB with a cosine potential, when the self-interaction increases or equivalently the decay constant decreases to about
$f \lesssim 10^{16}$ GeV, the star with the maximum compactness falls out of the LIGO most sensitive frequency band. On the other hand, if one only assumes the leading $|\phi|^4$ interaction,
a decay constant $f \sim 10^{16}$ GeV can still give rise to detectable boson star mergers.

\section{Conclusions and Outlook}
\label{sec:conclusion}
In this article, we focus on boson stars constituting light scalar
dark matter with repulsive self-interactions. We find spherically
symmetric ground state solutions for both the Schr\"odinger-Newton
equations and the full Klein-Gordon-Einstein equations without
approximations. Curiously there is no upper bound on the star mass
when using the former set of equations, while one finds a maximum mass
and compactness beyond which the star solutions are no longer stable
under perturbations solving the latter set of equations with weak
repulsion. We use a set of ansatz and give a heuristic argument for
the origin of the turning point between the stable and unstable
branches of solutions: for weak repulsion, the back-reaction of the
curvature on the scalar system can not be ignored, and can balance
against the self-repulsion.  We also point out that when one tries to
build a more realistic model of very light scalar dark matter in which
the mass is protected from quantum corrections by an approximate shift
symmetry, the scalar potential is usually significantly more
complicated than the widely used $|\phi|^4$ potential. For example, for
repulsive quartic interactions, such a potential will have
 cosine terms. Using the full potential, the
higher-order interactions are not negligible and the properties of the
boson star are very different, which also changes the prospect of GW
detection at interferometer experiments.

There are quite a few interesting open questions along this direction: 
\begin{itemize}
\item What is the maximum compactness of a boson star? Mergers of more compact stars have better detection prospects through GW probes at interferometer experiments such as LIGO. Assuming only the $|\phi|^4$ interaction, the maximum compactness of a spherically symmetric stable star is 0.16. In a more realistic pNGB model, the maximum compactness shrinks, however. It has been noted that for a double-well potential, the maximum compactness can be as large as 0.33~\cite{Bezares:2018qwa}. It would be worthwhile to explore whether there are particle physics models with protected small scalar mass and potential well approximated by the double-well potential. Another possibility is explore an excited star with a non-zero angular momentum~\cite{Sarkar:2017aje, Hertzberg:2018lmt}. 
\item What happens if the self-repulsion increases? In solving the Klein-Gordon-Einstein equation, we are restricted to relatively weak repulsion. When the repulsive self-interaction becomes strong, the system becomes highly non-linear and it is challenging to find stable solutions. It would be interesting to see whether there are qualitative changes in the profiles of the star solutions when the self-interaction strength increases. 
\item How do boson stars form? How do they evolve? We are agnostic about the dynamical evolution of the stars in the study. Will there be interesting signals from the collapse of unstable stars? 
\end{itemize}
Eventually we hope to map the macroscopic physics GW probes such as the mass and compactness of the merging object to the microscopic fundamental physics, such as the mass and decay constant of a light ALP.

\section*{Acknowledgment} We are grateful to Ewa Weinm\"uller for the
communications on numerical solutions to non-linear differential
equations. We also thank Robert Brandenberger, Rong-Gen Cai, Josh Eby, Marcelo
Gleiser, Thomas Helfer, Mark Hertzberg, David Kaiser, Masahiro Morii,
David Pinner, Matt Reece, Damian Sowinski, and Zhong-Zhi Xianyu for
useful discussions. JF is supported by the DOE grant DE-SC-0010010 and
NASA grant 80NSSC18K1010. CS is supported in part by the International
Postdoctoral Fellowship funded by China Postdoctoral Science
Foundation, and National Natural Science Foundation of China (No. 11875306). TRIUMF receives federal funding via a contribution
agreement with the National Research Council of Canada. JF and CS are
grateful for the hospitality of Boston University, where part of this work is completed.

\appendix

\subsection{A Real Scalar Model and Its Boson Star Solution}
\label{sec:cosine-potential-real}

\begin{figure}[t]
  \centering \includegraphics[width=.55\textwidth]{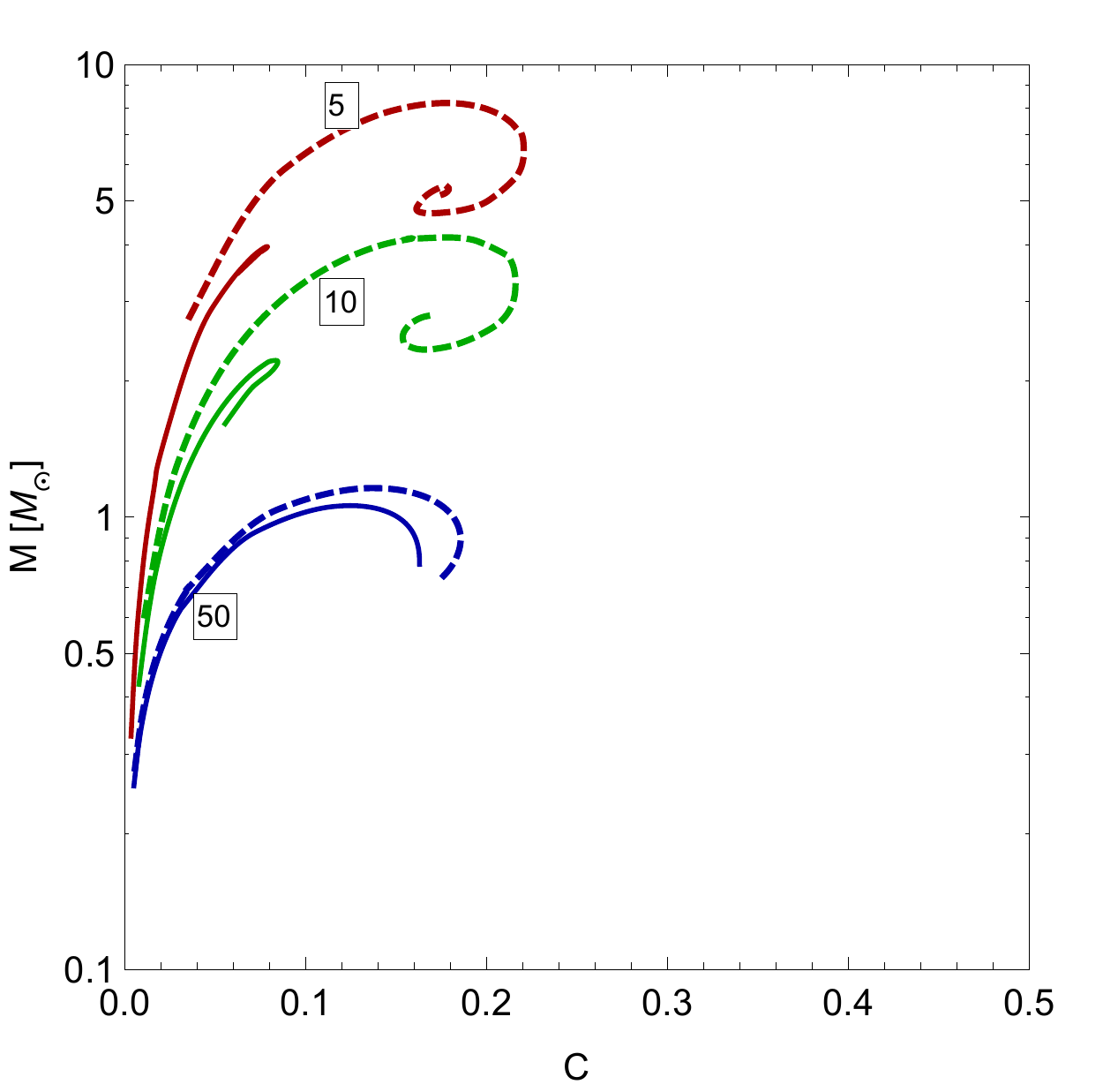
  }
  \caption{The compactness and mass profiles of boson stars in the
    pNGB model with a cosine potential (solid) and $\phi^4$ model
    (dashed) for three different $f$'s. All of the benchmark curves
    are generated with $m=10^{-10} \;\mathrm{eV}$.  Each curve is
    labeled by $f/(10^{16}\;\mathrm{GeV})$.  The cosine potential is
    in Eq.~\eqref{eq:cosine-potential-real} with
    $q_{F1} = 0.74, q_{F2} = 0.9, q_{B1} = 0.5, q_{B2} = 1$, and the
    $\phi^4$ interaction is chosen to match the leading term in the
    expansion of the cosine potential. 
  } 
  \label{fig:CM-phi4-cos-real}
\end{figure}

In this section, we will demonstrate that in a real scalar pNGB model, the boson star solution can be potentially very different using a full cosine potential, compared to using the leading $\phi^4$ truncation. 
One UV completed model of a real pNGB with leading self-interaction being repulsive is a $5D$ gauged $U(1)$ theory with the fifth dimension compactified on a circle. The pNGB is the gauge invariant Wilson loop of the fifth component of the gauge field. It obtains a one-loop effective potential from 5$D$ charged matter, e.g., $n_B$ bosons and $n_F$ fermions. We refer the readers to Ref.~\cite{Fan:2016rda} for details and only quote the $4D$ effective potential of the pNGB below
\be
  \label{eq:cosine-potential-real}
  V(\phi) \supset 
  \Lambda^4 \left ( \sum_{i=1}^{n_B} \cos \left ( \frac{q_{Bi}
  \phi}{f}\right ) -
  \sum_{i=1}^{n_F} \cos \left ( \frac{q_{Fi} \phi}{f} \right ) \right ),
\ee
where $q_{Bi}$ ($q_{Fi}$) is the charge of the $i$th boson
(fermion); $\Lambda$ and $f$ are two dimensionful parameters determined by the underlying $5D$ dynamics. To obtain the potential, we have assumed equal masses of all the charged matter for simplicity. 
Expanding the potential around the (local) minimum $\phi=0$, we have 
\bea
  \label{eq:cosine-exp-real}
  V(\phi)
  & =&
    \Lambda^4 \left [
    \frac{1}{2}\left(-\sum q_{Bi}^2 + \sum q_{Fi}^2\right) \frac{\phi^2}{f^2}+
    \frac{1}{4!}\left(\sum q_{B1}^4 -\sum q_{Fi}^4\right) \frac{\phi^4}{f^4} + ... 
    \right ]
    \cr
  & =&
    \frac{1}{2}m^2 \phi^2 + \frac{\lambda}{4!}\frac{m^2}{f^2}\phi^4
    +...    
\eea
with 
\bea
  m^2
  & =&
    Q_2 \left (\frac{\Lambda^4}{f^2} \right ), \cr
    \lambda \;
  & =&
     \frac{Q_4}{Q_2},
\eea
where $Q_2 =- \sum q_{Bi}^2 + \sum q_{Fi}^2$,
and $Q_4 = \sum q_{Bi}^4 - \sum q_{Fi}^4$.
The interaction size is controlled by the decay constant $f$, and the
sign is determined by the charge ratio $Q_4/Q_2$.
To get a positive mass and a repulsive interaction, the charge assignment has to satisfy
\bea
  \sum q_{Bi}^2 & <&   \sum q_{Fi}^2 , \nonumber \\
                  \sum q_{Bi}^4 & > & \sum q_{Fi}^4. 
   \label{eq:ineq-real}               
\eea
These inequalities are satisfied in some small corners of the charge space, for instance, $q_{F1} = 0.74, q_{F2} = 0.9, q_{B1} = 0.5, q_{B2} = 1$ in a model with two bosons and two fermions. Note that the quantized charges have to be integer multiples of a unit charge and the non-integer assignment here can be rescaled to be a set of integers.
Solving the equation of motion in GR numerically, we observe that the boson star solutions from the full potential are quite different from those from the truncation up to
$\phi^4$ in the mass profile. A few benchmark points are shown in Fig.~\ref{fig:CM-phi4-cos-real}. 

In finding the solution, we again use the single harmonic ansatz and  write the real scalar as a sum of two complex scalars $\phi (r,t)= (\psi (r,t) + \psi (r,t)^*)/\sqrt{2}$, where $\psi(r,t) = \psi(r) e^{-i m t}$. Note that
\bea
\int_{t_0}^{t_f} \; dt\;  \cos \phi
  & = &
\int_{t_0}^{t_f} \; dt\; \left (    \sum_{m=0}
    \frac{(-1)^m}{(2m)!}\phi^{2m}
    \right )
    \cr
  & = &
    \int_{t_0}^{t_f} \; dt\;   \left ( \sum_{m=0} \frac{(-1)^m}{(2m)!} C_{2m}^m \left ( \frac{ |
    {\psi}| }{\sqrt{2} } \right )^{2m} + (\text{fast mode})
    \right )
    \cr
  & \approx &
    \sum_{m=0} \frac{(-1)^m}{(m!)^2} \left (\frac{ \psi(r)}{\sqrt{2}}
    \right ) ^{2m}
     \times (t_f - t_0)
    \cr
  & = &
    J_0 (\sqrt{2}  {\psi}(r)) \times (t_f - t_0),
\eea
where $C_{2m}^m$ is the binomial coefficient and $J_0$ is the Bessel function of the first kind. This relation allows us to map (the time average of) a real
scalar theory with a cosine potential to a complex scalar theory with
a $J_0(\psi)$ potential, assuming we only take into account the
slow oscillation modes. {\hl As is known in the literature of boson
  stars, this approximation may lead to inaccurate results in the
  relativistic regime. It is still an open problem how to take into
  account of the fast oscillation modes in finding the boson star
  solutions numerically.}

\bibliography{bib}

\begin{thebibliography}{78}%
\makeatletter
\providecommand \@ifxundefined [1]{%
 \@ifx{#1\undefined}
}%
\providecommand \@ifnum [1]{%
 \ifnum #1\expandafter \@firstoftwo
 \else \expandafter \@secondoftwo
 \fi
}%
\providecommand \@ifx [1]{%
 \ifx #1\expandafter \@firstoftwo
 \else \expandafter \@secondoftwo
 \fi
}%
\providecommand \natexlab [1]{#1}%
\providecommand \enquote  [1]{``#1''}%
\providecommand \bibnamefont  [1]{#1}%
\providecommand \bibfnamefont [1]{#1}%
\providecommand \citenamefont [1]{#1}%
\providecommand \href@noop [0]{\@secondoftwo}%
\providecommand \href [0]{\begingroup \@sanitize@url \@href}%
\providecommand \@href[1]{\@@startlink{#1}\@@href}%
\providecommand \@@href[1]{\endgroup#1\@@endlink}%
\providecommand \@sanitize@url [0]{\catcode `\\12\catcode `\$12\catcode
  `\&12\catcode `\#12\catcode `\^12\catcode `\_12\catcode `\%12\relax}%
\providecommand \@@startlink[1]{}%
\providecommand \@@endlink[0]{}%
\providecommand \url  [0]{\begingroup\@sanitize@url \@url }%
\providecommand \@url [1]{\endgroup\@href {#1}{\urlprefix }}%
\providecommand \urlprefix  [0]{URL }%
\providecommand \Eprint [0]{\href }%
\providecommand \doibase [0]{http://dx.doi.org/}%
\providecommand \selectlanguage [0]{\@gobble}%
\providecommand \bibinfo  [0]{\@secondoftwo}%
\providecommand \bibfield  [0]{\@secondoftwo}%
\providecommand \translation [1]{[#1]}%
\providecommand \BibitemOpen [0]{}%
\providecommand \bibitemStop [0]{}%
\providecommand \bibitemNoStop [0]{.\EOS\space}%
\providecommand \EOS [0]{\spacefactor3000\relax}%
\providecommand \BibitemShut  [1]{\csname bibitem#1\endcsname}%
\let\auto@bib@innerbib\@empty
\bibitem [{\citenamefont {Abbott}\ \emph
  {et~al.}(2016{\natexlab{a}})\citenamefont {Abbott} \emph
  {et~al.}}]{TheLIGOScientific:2016agk}%
  \BibitemOpen
  \bibfield  {author} {\bibinfo {author} {\bibfnamefont {B.~P.}\ \bibnamefont
  {Abbott}} \emph {et~al.} (\bibinfo {collaboration} {Virgo, LIGO
  Scientific}),\ }\href {\doibase 10.1103/PhysRevLett.116.131103} {\bibfield
  {journal} {\bibinfo  {journal} {Phys. Rev. Lett.}\ }\textbf {\bibinfo
  {volume} {116}},\ \bibinfo {pages} {131103} (\bibinfo {year}
  {2016}{\natexlab{a}})},\ \Eprint {http://arxiv.org/abs/1602.03838}
  {arXiv:1602.03838 [gr-qc]} \BibitemShut {NoStop}%
\bibitem [{\citenamefont {Abbott}\ \emph
  {et~al.}(2016{\natexlab{b}})\citenamefont {Abbott} \emph
  {et~al.}}]{TheLIGOScientific:2016qqj}%
  \BibitemOpen
  \bibfield  {author} {\bibinfo {author} {\bibfnamefont {B.~P.}\ \bibnamefont
  {Abbott}} \emph {et~al.} (\bibinfo {collaboration} {Virgo, LIGO
  Scientific}),\ }\href {\doibase 10.1103/PhysRevD.93.122003} {\bibfield
  {journal} {\bibinfo  {journal} {Phys. Rev.}\ }\textbf {\bibinfo {volume}
  {D93}},\ \bibinfo {pages} {122003} (\bibinfo {year} {2016}{\natexlab{b}})},\
  \Eprint {http://arxiv.org/abs/1602.03839} {arXiv:1602.03839 [gr-qc]}
  \BibitemShut {NoStop}%
\bibitem [{\citenamefont {Abbott}\ \emph {et~al.}(2017)\citenamefont {Abbott}
  \emph {et~al.}}]{GBM:2017lvd}%
  \BibitemOpen
  \bibfield  {author} {\bibinfo {author} {\bibfnamefont {B.~P.}\ \bibnamefont
  {Abbott}} \emph {et~al.} (\bibinfo {collaboration} {GROND, SALT Group,
  OzGrav, DFN, INTEGRAL, Virgo, Insight-Hxmt, MAXI Team, Fermi-LAT, J-GEM,
  RATIR, IceCube, CAASTRO, LWA, ePESSTO, GRAWITA, RIMAS, SKA South
  Africa/MeerKAT, H.E.S.S., 1M2H Team, IKI-GW Follow-up, Fermi GBM, Pi of Sky,
  DWF (Deeper Wider Faster Program), Dark Energy Survey, MASTER, AstroSat
  Cadmium Zinc Telluride Imager Team, Swift, Pierre Auger, ASKAP, VINROUGE,
  JAGWAR, Chandra Team at McGill University, TTU-NRAO, GROWTH, AGILE Team, MWA,
  ATCA, AST3, TOROS, Pan-STARRS, NuSTAR, ATLAS Telescopes, BOOTES, CaltechNRAO,
  LIGO Scientific, High Time Resolution Universe Survey, Nordic Optical
  Telescope, Las Cumbres Observatory Group, TZAC Consortium, LOFAR, IPN, DLT40,
  Texas Tech University, HAWC, ANTARES, KU, Dark Energy Camera GW-EM, CALET,
  Euro VLBI Team, ALMA}),\ }\href {\doibase 10.3847/2041-8213/aa91c9}
  {\bibfield  {journal} {\bibinfo  {journal} {Astrophys. J.}\ }\textbf
  {\bibinfo {volume} {848}},\ \bibinfo {pages} {L12} (\bibinfo {year}
  {2017})},\ \Eprint {http://arxiv.org/abs/1710.05833} {arXiv:1710.05833
  [astro-ph.HE]} \BibitemShut {NoStop}%
\bibitem [{\citenamefont {Giudice}\ \emph {et~al.}(2016)\citenamefont
  {Giudice}, \citenamefont {McCullough},\ and\ \citenamefont
  {Urbano}}]{Giudice:2016zpa}%
  \BibitemOpen
  \bibfield  {author} {\bibinfo {author} {\bibfnamefont {G.~F.}\ \bibnamefont
  {Giudice}}, \bibinfo {author} {\bibfnamefont {M.}~\bibnamefont {McCullough}},
  \ and\ \bibinfo {author} {\bibfnamefont {A.}~\bibnamefont {Urbano}},\ }\href
  {\doibase 10.1088/1475-7516/2016/10/001} {\bibfield  {journal} {\bibinfo
  {journal} {JCAP}\ }\textbf {\bibinfo {volume} {1610}},\ \bibinfo {pages}
  {001} (\bibinfo {year} {2016})},\ \Eprint {http://arxiv.org/abs/1605.01209}
  {arXiv:1605.01209 [hep-ph]} \BibitemShut {NoStop}%
\bibitem [{\citenamefont {Dev}\ \emph {et~al.}(2017)\citenamefont {Dev},
  \citenamefont {Lindner},\ and\ \citenamefont {Ohmer}}]{Dev:2016hxv}%
  \BibitemOpen
  \bibfield  {author} {\bibinfo {author} {\bibfnamefont {P.~S.~B.}\
  \bibnamefont {Dev}}, \bibinfo {author} {\bibfnamefont {M.}~\bibnamefont
  {Lindner}}, \ and\ \bibinfo {author} {\bibfnamefont {S.}~\bibnamefont
  {Ohmer}},\ }\href {\doibase 10.1016/j.physletb.2017.08.043} {\bibfield
  {journal} {\bibinfo  {journal} {Phys. Lett.}\ }\textbf {\bibinfo {volume}
  {B773}},\ \bibinfo {pages} {219} (\bibinfo {year} {2017})},\ \Eprint
  {http://arxiv.org/abs/1609.03939} {arXiv:1609.03939 [hep-ph]} \BibitemShut
  {NoStop}%
\bibitem [{\citenamefont {Ellis}\ \emph {et~al.}(2018)\citenamefont {Ellis},
  \citenamefont {Hektor}, \citenamefont {Hutsi}, \citenamefont {Kannike},
  \citenamefont {Marzola}, \citenamefont {Raidal},\ and\ \citenamefont
  {Vaskonen}}]{Ellis:2017jgp}%
  \BibitemOpen
  \bibfield  {author} {\bibinfo {author} {\bibfnamefont {J.}~\bibnamefont
  {Ellis}}, \bibinfo {author} {\bibfnamefont {A.}~\bibnamefont {Hektor}},
  \bibinfo {author} {\bibfnamefont {G.}~\bibnamefont {Hutsi}}, \bibinfo
  {author} {\bibfnamefont {K.}~\bibnamefont {Kannike}}, \bibinfo {author}
  {\bibfnamefont {L.}~\bibnamefont {Marzola}}, \bibinfo {author} {\bibfnamefont
  {M.}~\bibnamefont {Raidal}}, \ and\ \bibinfo {author} {\bibfnamefont
  {V.}~\bibnamefont {Vaskonen}},\ }\href {\doibase
  10.1016/j.physletb.2018.04.048} {\bibfield  {journal} {\bibinfo  {journal}
  {Phys. Lett.}\ }\textbf {\bibinfo {volume} {B781}},\ \bibinfo {pages} {607}
  (\bibinfo {year} {2018})},\ \Eprint {http://arxiv.org/abs/1710.05540}
  {arXiv:1710.05540 [astro-ph.CO]} \BibitemShut {NoStop}%
\bibitem [{\citenamefont {Croon}\ \emph
  {et~al.}(2018{\natexlab{a}})\citenamefont {Croon}, \citenamefont {Nelson},
  \citenamefont {Sun}, \citenamefont {Walker},\ and\ \citenamefont
  {Xianyu}}]{Croon:2017zcu}%
  \BibitemOpen
  \bibfield  {author} {\bibinfo {author} {\bibfnamefont {D.}~\bibnamefont
  {Croon}}, \bibinfo {author} {\bibfnamefont {A.~E.}\ \bibnamefont {Nelson}},
  \bibinfo {author} {\bibfnamefont {C.}~\bibnamefont {Sun}}, \bibinfo {author}
  {\bibfnamefont {D.~G.~E.}\ \bibnamefont {Walker}}, \ and\ \bibinfo {author}
  {\bibfnamefont {Z.-Z.}\ \bibnamefont {Xianyu}},\ }\href {\doibase
  10.3847/2041-8213/aabe76} {\bibfield  {journal} {\bibinfo  {journal}
  {Astrophys. J.}\ }\textbf {\bibinfo {volume} {858}},\ \bibinfo {pages} {L2}
  (\bibinfo {year} {2018}{\natexlab{a}})},\ \Eprint
  {http://arxiv.org/abs/1711.02096} {arXiv:1711.02096 [hep-ph]} \BibitemShut
  {NoStop}%
\bibitem [{\citenamefont {Randall}\ and\ \citenamefont
  {Xianyu}(2018)}]{Randall:2017jop}%
  \BibitemOpen
  \bibfield  {author} {\bibinfo {author} {\bibfnamefont {L.}~\bibnamefont
  {Randall}}\ and\ \bibinfo {author} {\bibfnamefont {Z.-Z.}\ \bibnamefont
  {Xianyu}},\ }\href {\doibase 10.3847/1538-4357/aaa1a2} {\bibfield  {journal}
  {\bibinfo  {journal} {Astrophys. J.}\ }\textbf {\bibinfo {volume} {853}},\
  \bibinfo {pages} {93} (\bibinfo {year} {2018})},\ \Eprint
  {http://arxiv.org/abs/1708.08569} {arXiv:1708.08569 [gr-qc]} \BibitemShut
  {NoStop}%
\bibitem [{\citenamefont {Cui}\ \emph {et~al.}(2018)\citenamefont {Cui},
  \citenamefont {Lewicki}, \citenamefont {Morrissey},\ and\ \citenamefont
  {Wells}}]{Cui:2017ufi}%
  \BibitemOpen
  \bibfield  {author} {\bibinfo {author} {\bibfnamefont {Y.}~\bibnamefont
  {Cui}}, \bibinfo {author} {\bibfnamefont {M.}~\bibnamefont {Lewicki}},
  \bibinfo {author} {\bibfnamefont {D.~E.}\ \bibnamefont {Morrissey}}, \ and\
  \bibinfo {author} {\bibfnamefont {J.~D.}\ \bibnamefont {Wells}},\ }\href
  {\doibase 10.1103/PhysRevD.97.123505} {\bibfield  {journal} {\bibinfo
  {journal} {Phys. Rev.}\ }\textbf {\bibinfo {volume} {D97}},\ \bibinfo {pages}
  {123505} (\bibinfo {year} {2018})},\ \Eprint
  {http://arxiv.org/abs/1711.03104} {arXiv:1711.03104 [hep-ph]} \BibitemShut
  {NoStop}%
\bibitem [{\citenamefont {Pierce}\ \emph {et~al.}(2018)\citenamefont {Pierce},
  \citenamefont {Riles},\ and\ \citenamefont {Zhao}}]{Pierce:2018xmy}%
  \BibitemOpen
  \bibfield  {author} {\bibinfo {author} {\bibfnamefont {A.}~\bibnamefont
  {Pierce}}, \bibinfo {author} {\bibfnamefont {K.}~\bibnamefont {Riles}}, \
  and\ \bibinfo {author} {\bibfnamefont {Y.}~\bibnamefont {Zhao}},\ }\href
  {\doibase 10.1103/PhysRevLett.121.061102} {\bibfield  {journal} {\bibinfo
  {journal} {Phys. Rev. Lett.}\ }\textbf {\bibinfo {volume} {121}},\ \bibinfo
  {pages} {061102} (\bibinfo {year} {2018})},\ \Eprint
  {http://arxiv.org/abs/1801.10161} {arXiv:1801.10161 [hep-ph]} \BibitemShut
  {NoStop}%
\bibitem [{\citenamefont {Amin}\ \emph {et~al.}(2018)\citenamefont {Amin},
  \citenamefont {Fan}, \citenamefont {Lozanov},\ and\ \citenamefont
  {Reece}}]{Amin:2018kkg}%
  \BibitemOpen
  \bibfield  {author} {\bibinfo {author} {\bibfnamefont {M.~A.}\ \bibnamefont
  {Amin}}, \bibinfo {author} {\bibfnamefont {J.}~\bibnamefont {Fan}}, \bibinfo
  {author} {\bibfnamefont {K.~D.}\ \bibnamefont {Lozanov}}, \ and\ \bibinfo
  {author} {\bibfnamefont {M.}~\bibnamefont {Reece}},\ }\href@noop {} {\
  (\bibinfo {year} {2018})},\ \Eprint {http://arxiv.org/abs/1802.00444}
  {arXiv:1802.00444 [hep-ph]} \BibitemShut {NoStop}%
\bibitem [{\citenamefont {Nelson}\ \emph {et~al.}(2018)\citenamefont {Nelson},
  \citenamefont {Reddy},\ and\ \citenamefont {Zhou}}]{Nelson:2018xtr}%
  \BibitemOpen
  \bibfield  {author} {\bibinfo {author} {\bibfnamefont {A.}~\bibnamefont
  {Nelson}}, \bibinfo {author} {\bibfnamefont {S.}~\bibnamefont {Reddy}}, \
  and\ \bibinfo {author} {\bibfnamefont {D.}~\bibnamefont {Zhou}},\ }\href@noop
  {} {\  (\bibinfo {year} {2018})},\ \Eprint {http://arxiv.org/abs/1803.03266}
  {arXiv:1803.03266 [hep-ph]} \BibitemShut {NoStop}%
\bibitem [{\citenamefont {Geller}\ \emph {et~al.}(2018)\citenamefont {Geller},
  \citenamefont {Hook}, \citenamefont {Sundrum},\ and\ \citenamefont
  {Tsai}}]{Geller:2018mwu}%
  \BibitemOpen
  \bibfield  {author} {\bibinfo {author} {\bibfnamefont {M.}~\bibnamefont
  {Geller}}, \bibinfo {author} {\bibfnamefont {A.}~\bibnamefont {Hook}},
  \bibinfo {author} {\bibfnamefont {R.}~\bibnamefont {Sundrum}}, \ and\
  \bibinfo {author} {\bibfnamefont {Y.}~\bibnamefont {Tsai}},\ }\href@noop {}
  {\  (\bibinfo {year} {2018})},\ \Eprint {http://arxiv.org/abs/1803.10780}
  {arXiv:1803.10780 [hep-ph]} \BibitemShut {NoStop}%
\bibitem [{\citenamefont {Croon}\ \emph
  {et~al.}(2018{\natexlab{b}})\citenamefont {Croon}, \citenamefont {Sanz},\
  and\ \citenamefont {White}}]{Croon:2018erz}%
  \BibitemOpen
  \bibfield  {author} {\bibinfo {author} {\bibfnamefont {D.}~\bibnamefont
  {Croon}}, \bibinfo {author} {\bibfnamefont {V.}~\bibnamefont {Sanz}}, \ and\
  \bibinfo {author} {\bibfnamefont {G.}~\bibnamefont {White}},\ }\href
  {\doibase 10.1007/JHEP08(2018)203} {\bibfield  {journal} {\bibinfo  {journal}
  {JHEP}\ }\textbf {\bibinfo {volume} {08}},\ \bibinfo {pages} {203} (\bibinfo
  {year} {2018}{\natexlab{b}})},\ \Eprint {http://arxiv.org/abs/1806.02332}
  {arXiv:1806.02332 [hep-ph]} \BibitemShut {NoStop}%
\bibitem [{\citenamefont {Figueroa}\ \emph {et~al.}(2018)\citenamefont
  {Figueroa}, \citenamefont {Megias}, \citenamefont {Nardini}, \citenamefont
  {Pieroni}, \citenamefont {Quiros}, \citenamefont {Ricciardone},\ and\
  \citenamefont {Tasinato}}]{Figueroa:2018xtu}%
  \BibitemOpen
  \bibfield  {author} {\bibinfo {author} {\bibfnamefont {D.~G.}\ \bibnamefont
  {Figueroa}}, \bibinfo {author} {\bibfnamefont {E.}~\bibnamefont {Megias}},
  \bibinfo {author} {\bibfnamefont {G.}~\bibnamefont {Nardini}}, \bibinfo
  {author} {\bibfnamefont {M.}~\bibnamefont {Pieroni}}, \bibinfo {author}
  {\bibfnamefont {M.}~\bibnamefont {Quiros}}, \bibinfo {author} {\bibfnamefont
  {A.}~\bibnamefont {Ricciardone}}, \ and\ \bibinfo {author} {\bibfnamefont
  {G.}~\bibnamefont {Tasinato}},\ }\href@noop {} {\  (\bibinfo {year}
  {2018})},\ \Eprint {http://arxiv.org/abs/1806.06463} {arXiv:1806.06463
  [astro-ph.CO]} \BibitemShut {NoStop}%
\bibitem [{\citenamefont {Bauswein}\ \emph {et~al.}(2018)\citenamefont
  {Bauswein}, \citenamefont {Bastian}, \citenamefont {Blaschke}, \citenamefont
  {Chatziioannou}, \citenamefont {Clark}, \citenamefont {Fischer},\ and\
  \citenamefont {Oertel}}]{Bauswein:2018bma}%
  \BibitemOpen
  \bibfield  {author} {\bibinfo {author} {\bibfnamefont {A.}~\bibnamefont
  {Bauswein}}, \bibinfo {author} {\bibfnamefont {N.-U.~F.}\ \bibnamefont
  {Bastian}}, \bibinfo {author} {\bibfnamefont {D.~B.}\ \bibnamefont
  {Blaschke}}, \bibinfo {author} {\bibfnamefont {K.}~\bibnamefont
  {Chatziioannou}}, \bibinfo {author} {\bibfnamefont {J.~A.}\ \bibnamefont
  {Clark}}, \bibinfo {author} {\bibfnamefont {T.}~\bibnamefont {Fischer}}, \
  and\ \bibinfo {author} {\bibfnamefont {M.}~\bibnamefont {Oertel}},\
  }\href@noop {} {\  (\bibinfo {year} {2018})},\ \Eprint
  {http://arxiv.org/abs/1809.01116} {arXiv:1809.01116 [astro-ph.HE]}
  \BibitemShut {NoStop}%
\bibitem [{\citenamefont {Kopp}\ \emph {et~al.}(2018)\citenamefont {Kopp},
  \citenamefont {Laha}, \citenamefont {Opferkuch},\ and\ \citenamefont
  {Shepherd}}]{Kopp:2018jom}%
  \BibitemOpen
  \bibfield  {author} {\bibinfo {author} {\bibfnamefont {J.}~\bibnamefont
  {Kopp}}, \bibinfo {author} {\bibfnamefont {R.}~\bibnamefont {Laha}}, \bibinfo
  {author} {\bibfnamefont {T.}~\bibnamefont {Opferkuch}}, \ and\ \bibinfo
  {author} {\bibfnamefont {W.}~\bibnamefont {Shepherd}},\ }\href {\doibase
  10.1007/JHEP11(2018)096} {\bibfield  {journal} {\bibinfo  {journal} {JHEP}\
  }\textbf {\bibinfo {volume} {11}},\ \bibinfo {pages} {096} (\bibinfo {year}
  {2018})},\ \Eprint {http://arxiv.org/abs/1807.02527} {arXiv:1807.02527
  [hep-ph]} \BibitemShut {NoStop}%
\bibitem [{\citenamefont {Alexander}\ \emph {et~al.}(2018)\citenamefont
  {Alexander}, \citenamefont {McDonough}, \citenamefont {Sims},\ and\
  \citenamefont {Yunes}}]{Alexander:2018qzg}%
  \BibitemOpen
  \bibfield  {author} {\bibinfo {author} {\bibfnamefont {S.}~\bibnamefont
  {Alexander}}, \bibinfo {author} {\bibfnamefont {E.}~\bibnamefont
  {McDonough}}, \bibinfo {author} {\bibfnamefont {R.}~\bibnamefont {Sims}}, \
  and\ \bibinfo {author} {\bibfnamefont {N.}~\bibnamefont {Yunes}},\ }\href
  {\doibase 10.1088/1361-6382/aaeb5c} {\bibfield  {journal} {\bibinfo
  {journal} {Class. Quant. Grav.}\ }\textbf {\bibinfo {volume} {35}},\ \bibinfo
  {pages} {235012} (\bibinfo {year} {2018})},\ \Eprint
  {http://arxiv.org/abs/1808.05286} {arXiv:1808.05286 [gr-qc]} \BibitemShut
  {NoStop}%
\bibitem [{\citenamefont {Tkachev}(1986)}]{Tkachev:1986tr}%
  \BibitemOpen
  \bibfield  {author} {\bibinfo {author} {\bibfnamefont {I.~I.}\ \bibnamefont
  {Tkachev}},\ }\href@noop {} {\bibfield  {journal} {\bibinfo  {journal} {Sov.
  Astron. Lett.}\ }\textbf {\bibinfo {volume} {12}},\ \bibinfo {pages} {305}
  (\bibinfo {year} {1986})},\ \bibinfo {note} {[Pisma Astron.
  Zh.12,726(1986)]}\BibitemShut {NoStop}%
\bibitem [{\citenamefont {Peccei}\ and\ \citenamefont
  {Quinn}(1977{\natexlab{a}})}]{Peccei:1977ur}%
  \BibitemOpen
  \bibfield  {author} {\bibinfo {author} {\bibfnamefont {R.~D.}\ \bibnamefont
  {Peccei}}\ and\ \bibinfo {author} {\bibfnamefont {H.~R.}\ \bibnamefont
  {Quinn}},\ }\href {\doibase 10.1103/PhysRevD.16.1791} {\bibfield  {journal}
  {\bibinfo  {journal} {Phys. Rev.}\ }\textbf {\bibinfo {volume} {D16}},\
  \bibinfo {pages} {1791} (\bibinfo {year} {1977}{\natexlab{a}})}\BibitemShut
  {NoStop}%
\bibitem [{\citenamefont {Peccei}\ and\ \citenamefont
  {Quinn}(1977{\natexlab{b}})}]{Peccei:1977hh}%
  \BibitemOpen
  \bibfield  {author} {\bibinfo {author} {\bibfnamefont {R.~D.}\ \bibnamefont
  {Peccei}}\ and\ \bibinfo {author} {\bibfnamefont {H.~R.}\ \bibnamefont
  {Quinn}},\ }\href {\doibase 10.1103/PhysRevLett.38.1440} {\bibfield
  {journal} {\bibinfo  {journal} {Phys. Rev. Lett.}\ }\textbf {\bibinfo
  {volume} {38}},\ \bibinfo {pages} {1440} (\bibinfo {year}
  {1977}{\natexlab{b}})}\BibitemShut {NoStop}%
\bibitem [{\citenamefont {Wilczek}(1978)}]{Wilczek:1977pj}%
  \BibitemOpen
  \bibfield  {author} {\bibinfo {author} {\bibfnamefont {F.}~\bibnamefont
  {Wilczek}},\ }\href {\doibase 10.1103/PhysRevLett.40.279} {\bibfield
  {journal} {\bibinfo  {journal} {Phys. Rev. Lett.}\ }\textbf {\bibinfo
  {volume} {40}},\ \bibinfo {pages} {279} (\bibinfo {year} {1978})}\BibitemShut
  {NoStop}%
\bibitem [{\citenamefont {Weinberg}(1978)}]{Weinberg:1977ma}%
  \BibitemOpen
  \bibfield  {author} {\bibinfo {author} {\bibfnamefont {S.}~\bibnamefont
  {Weinberg}},\ }\href {\doibase 10.1103/PhysRevLett.40.223} {\bibfield
  {journal} {\bibinfo  {journal} {Phys. Rev. Lett.}\ }\textbf {\bibinfo
  {volume} {40}},\ \bibinfo {pages} {223} (\bibinfo {year} {1978})}\BibitemShut
  {NoStop}%
\bibitem [{\citenamefont {Kim}(1979)}]{Kim:1979if}%
  \BibitemOpen
  \bibfield  {author} {\bibinfo {author} {\bibfnamefont {J.~E.}\ \bibnamefont
  {Kim}},\ }\href {\doibase 10.1103/PhysRevLett.43.103} {\bibfield  {journal}
  {\bibinfo  {journal} {Phys. Rev. Lett.}\ }\textbf {\bibinfo {volume} {43}},\
  \bibinfo {pages} {103} (\bibinfo {year} {1979})}\BibitemShut {NoStop}%
\bibitem [{\citenamefont {Shifman}\ \emph {et~al.}(1980)\citenamefont
  {Shifman}, \citenamefont {Vainshtein},\ and\ \citenamefont
  {Zakharov}}]{Shifman:1979if}%
  \BibitemOpen
  \bibfield  {author} {\bibinfo {author} {\bibfnamefont {M.~A.}\ \bibnamefont
  {Shifman}}, \bibinfo {author} {\bibfnamefont {A.~I.}\ \bibnamefont
  {Vainshtein}}, \ and\ \bibinfo {author} {\bibfnamefont {V.~I.}\ \bibnamefont
  {Zakharov}},\ }\href {\doibase 10.1016/0550-3213(80)90209-6} {\bibfield
  {journal} {\bibinfo  {journal} {Nucl. Phys.}\ }\textbf {\bibinfo {volume}
  {B166}},\ \bibinfo {pages} {493} (\bibinfo {year} {1980})}\BibitemShut
  {NoStop}%
\bibitem [{\citenamefont {Zhitnitsky}(1980)}]{Zhitnitsky:1980tq}%
  \BibitemOpen
  \bibfield  {author} {\bibinfo {author} {\bibfnamefont {A.~R.}\ \bibnamefont
  {Zhitnitsky}},\ }\href@noop {} {\bibfield  {journal} {\bibinfo  {journal}
  {Sov. J. Nucl. Phys.}\ }\textbf {\bibinfo {volume} {31}},\ \bibinfo {pages}
  {260} (\bibinfo {year} {1980})},\ \bibinfo {note} {[Yad.
  Fiz.31,497(1980)]}\BibitemShut {NoStop}%
\bibitem [{\citenamefont {Dine}\ \emph {et~al.}(1981)\citenamefont {Dine},
  \citenamefont {Fischler},\ and\ \citenamefont {Srednicki}}]{Dine:1981rt}%
  \BibitemOpen
  \bibfield  {author} {\bibinfo {author} {\bibfnamefont {M.}~\bibnamefont
  {Dine}}, \bibinfo {author} {\bibfnamefont {W.}~\bibnamefont {Fischler}}, \
  and\ \bibinfo {author} {\bibfnamefont {M.}~\bibnamefont {Srednicki}},\ }\href
  {\doibase 10.1016/0370-2693(81)90590-6} {\bibfield  {journal} {\bibinfo
  {journal} {Phys. Lett.}\ }\textbf {\bibinfo {volume} {104B}},\ \bibinfo
  {pages} {199} (\bibinfo {year} {1981})}\BibitemShut {NoStop}%
\bibitem [{\citenamefont {Visinelli}\ \emph {et~al.}(2018)\citenamefont
  {Visinelli}, \citenamefont {Baum}, \citenamefont {Redondo}, \citenamefont
  {Freese},\ and\ \citenamefont {Wilczek}}]{Visinelli:2017ooc}%
  \BibitemOpen
  \bibfield  {author} {\bibinfo {author} {\bibfnamefont {L.}~\bibnamefont
  {Visinelli}}, \bibinfo {author} {\bibfnamefont {S.}~\bibnamefont {Baum}},
  \bibinfo {author} {\bibfnamefont {J.}~\bibnamefont {Redondo}}, \bibinfo
  {author} {\bibfnamefont {K.}~\bibnamefont {Freese}}, \ and\ \bibinfo {author}
  {\bibfnamefont {F.}~\bibnamefont {Wilczek}},\ }\href {\doibase
  10.1016/j.physletb.2017.12.010} {\bibfield  {journal} {\bibinfo  {journal}
  {Phys. Lett.}\ }\textbf {\bibinfo {volume} {B777}},\ \bibinfo {pages} {64}
  (\bibinfo {year} {2018})},\ \Eprint {http://arxiv.org/abs/1710.08910}
  {arXiv:1710.08910 [astro-ph.CO]} \BibitemShut {NoStop}%
\bibitem [{\citenamefont {Schiappacasse}\ and\ \citenamefont
  {Hertzberg}(2018)}]{Schiappacasse:2017ham}%
  \BibitemOpen
  \bibfield  {author} {\bibinfo {author} {\bibfnamefont {E.~D.}\ \bibnamefont
  {Schiappacasse}}\ and\ \bibinfo {author} {\bibfnamefont {M.~P.}\ \bibnamefont
  {Hertzberg}},\ }\href {\doibase 10.1088/1475-7516/2018/03/E01,
  10.1088/1475-7516/2018/01/037} {\bibfield  {journal} {\bibinfo  {journal}
  {JCAP}\ }\textbf {\bibinfo {volume} {1801}},\ \bibinfo {pages} {037}
  (\bibinfo {year} {2018})},\ \bibinfo {note} {[Erratum:
  JCAP1803,no.03,E01(2018)]},\ \Eprint {http://arxiv.org/abs/1710.04729}
  {arXiv:1710.04729 [hep-ph]} \BibitemShut {NoStop}%
\bibitem [{\citenamefont {Chavanis}(2018)}]{Chavanis:2017loo}%
  \BibitemOpen
  \bibfield  {author} {\bibinfo {author} {\bibfnamefont {P.-H.}\ \bibnamefont
  {Chavanis}},\ }\href {\doibase 10.1103/PhysRevD.98.023009} {\bibfield
  {journal} {\bibinfo  {journal} {Phys. Rev.}\ }\textbf {\bibinfo {volume}
  {D98}},\ \bibinfo {pages} {023009} (\bibinfo {year} {2018})},\ \Eprint
  {http://arxiv.org/abs/1710.06268} {arXiv:1710.06268 [gr-qc]} \BibitemShut
  {NoStop}%
\bibitem [{\citenamefont {Eby}\ \emph {et~al.}(2018{\natexlab{a}})\citenamefont
  {Eby}, \citenamefont {Suranyi},\ and\ \citenamefont
  {Wijewardhana}}]{Eby:2017teq}%
  \BibitemOpen
  \bibfield  {author} {\bibinfo {author} {\bibfnamefont {J.}~\bibnamefont
  {Eby}}, \bibinfo {author} {\bibfnamefont {P.}~\bibnamefont {Suranyi}}, \ and\
  \bibinfo {author} {\bibfnamefont {L.~C.~R.}\ \bibnamefont {Wijewardhana}},\
  }\href {\doibase 10.1088/1475-7516/2018/04/038} {\bibfield  {journal}
  {\bibinfo  {journal} {JCAP}\ }\textbf {\bibinfo {volume} {1804}},\ \bibinfo
  {pages} {038} (\bibinfo {year} {2018}{\natexlab{a}})},\ \Eprint
  {http://arxiv.org/abs/1712.04941} {arXiv:1712.04941 [hep-ph]} \BibitemShut
  {NoStop}%
\bibitem [{\citenamefont {Goodman}(2000)}]{Goodman:2000tg}%
  \BibitemOpen
  \bibfield  {author} {\bibinfo {author} {\bibfnamefont {J.}~\bibnamefont
  {Goodman}},\ }\href {\doibase 10.1016/S1384-1076(00)00015-4} {\bibfield
  {journal} {\bibinfo  {journal} {New Astron.}\ }\textbf {\bibinfo {volume}
  {5}},\ \bibinfo {pages} {103} (\bibinfo {year} {2000})},\ \Eprint
  {http://arxiv.org/abs/astro-ph/0003018} {arXiv:astro-ph/0003018 [astro-ph]}
  \BibitemShut {NoStop}%
\bibitem [{\citenamefont {Peebles}(2000)}]{Peebles:2000yy}%
  \BibitemOpen
  \bibfield  {author} {\bibinfo {author} {\bibfnamefont {P.~J.~E.}\
  \bibnamefont {Peebles}},\ }\href {\doibase 10.1086/312677} {\bibfield
  {journal} {\bibinfo  {journal} {Astrophys. J.}\ }\textbf {\bibinfo {volume}
  {534}},\ \bibinfo {pages} {L127} (\bibinfo {year} {2000})},\ \Eprint
  {http://arxiv.org/abs/astro-ph/0002495} {arXiv:astro-ph/0002495 [astro-ph]}
  \BibitemShut {NoStop}%
\bibitem [{\citenamefont {Flores}\ and\ \citenamefont
  {Primack}(1994)}]{Flores:1994gz}%
  \BibitemOpen
  \bibfield  {author} {\bibinfo {author} {\bibfnamefont {R.~A.}\ \bibnamefont
  {Flores}}\ and\ \bibinfo {author} {\bibfnamefont {J.~R.}\ \bibnamefont
  {Primack}},\ }\href {\doibase 10.1086/187350} {\bibfield  {journal} {\bibinfo
   {journal} {Astrophys. J.}\ }\textbf {\bibinfo {volume} {427}},\ \bibinfo
  {pages} {L1} (\bibinfo {year} {1994})},\ \Eprint
  {http://arxiv.org/abs/astro-ph/9402004} {arXiv:astro-ph/9402004 [astro-ph]}
  \BibitemShut {NoStop}%
\bibitem [{\citenamefont {{Moore}}(1994)}]{1994Natur.370..629M}%
  \BibitemOpen
  \bibfield  {author} {\bibinfo {author} {\bibfnamefont {B.}~\bibnamefont
  {{Moore}}},\ }\href {\doibase 10.1038/370629a0} {\bibfield  {journal}
  {\bibinfo  {journal} {Nature}\ }\textbf {\bibinfo {volume} {370}},\ \bibinfo
  {pages} {629} (\bibinfo {year} {1994})}\BibitemShut {NoStop}%
\bibitem [{\citenamefont {Burkert}(1996)}]{Burkert:1995yz}%
  \BibitemOpen
  \bibfield  {author} {\bibinfo {author} {\bibfnamefont {A.}~\bibnamefont
  {Burkert}},\ }\bibfield  {booktitle} {\emph {\bibinfo {booktitle} {{IAU
  Symposium 171: New Light on Galaxy Evolution Heidelberg, Germany, June 26-30,
  1995}}},\ }\href {\doibase 10.1086/309560} {\bibfield  {journal} {\bibinfo
  {journal} {IAU Symp.}\ }\textbf {\bibinfo {volume} {171}},\ \bibinfo {pages}
  {175} (\bibinfo {year} {1996})},\ \bibinfo {note} {[Astrophys.
  J.447,L25(1995)]},\ \Eprint {http://arxiv.org/abs/astro-ph/9504041}
  {arXiv:astro-ph/9504041 [astro-ph]} \BibitemShut {NoStop}%
\bibitem [{\citenamefont {Moore}\ \emph {et~al.}(1999)\citenamefont {Moore},
  \citenamefont {Quinn}, \citenamefont {Governato}, \citenamefont {Stadel},\
  and\ \citenamefont {Lake}}]{Moore:1999gc}%
  \BibitemOpen
  \bibfield  {author} {\bibinfo {author} {\bibfnamefont {B.}~\bibnamefont
  {Moore}}, \bibinfo {author} {\bibfnamefont {T.~R.}\ \bibnamefont {Quinn}},
  \bibinfo {author} {\bibfnamefont {F.}~\bibnamefont {Governato}}, \bibinfo
  {author} {\bibfnamefont {J.}~\bibnamefont {Stadel}}, \ and\ \bibinfo {author}
  {\bibfnamefont {G.}~\bibnamefont {Lake}},\ }\href {\doibase
  10.1046/j.1365-8711.1999.03039.x} {\bibfield  {journal} {\bibinfo  {journal}
  {Mon. Not. Roy. Astron. Soc.}\ }\textbf {\bibinfo {volume} {310}},\ \bibinfo
  {pages} {1147} (\bibinfo {year} {1999})},\ \Eprint
  {http://arxiv.org/abs/astro-ph/9903164} {arXiv:astro-ph/9903164 [astro-ph]}
  \BibitemShut {NoStop}%
\bibitem [{\citenamefont {Salucci}\ and\ \citenamefont
  {Burkert}(2000)}]{Salucci:2000ps}%
  \BibitemOpen
  \bibfield  {author} {\bibinfo {author} {\bibfnamefont {P.}~\bibnamefont
  {Salucci}}\ and\ \bibinfo {author} {\bibfnamefont {A.}~\bibnamefont
  {Burkert}},\ }\href {\doibase 10.1086/312747} {\bibfield  {journal} {\bibinfo
   {journal} {Astrophys. J.}\ }\textbf {\bibinfo {volume} {537}},\ \bibinfo
  {pages} {L9} (\bibinfo {year} {2000})},\ \Eprint
  {http://arxiv.org/abs/astro-ph/0004397} {arXiv:astro-ph/0004397 [astro-ph]}
  \BibitemShut {NoStop}%
\bibitem [{\citenamefont {Deng}\ \emph {et~al.}(2018)\citenamefont {Deng},
  \citenamefont {Hertzberg}, \citenamefont {Namjoo},\ and\ \citenamefont
  {Masoumi}}]{Deng:2018jjz}%
  \BibitemOpen
  \bibfield  {author} {\bibinfo {author} {\bibfnamefont {H.}~\bibnamefont
  {Deng}}, \bibinfo {author} {\bibfnamefont {M.~P.}\ \bibnamefont {Hertzberg}},
  \bibinfo {author} {\bibfnamefont {M.~H.}\ \bibnamefont {Namjoo}}, \ and\
  \bibinfo {author} {\bibfnamefont {A.}~\bibnamefont {Masoumi}},\ }\href
  {\doibase 10.1103/PhysRevD.98.023513} {\bibfield  {journal} {\bibinfo
  {journal} {Phys. Rev.}\ }\textbf {\bibinfo {volume} {D98}},\ \bibinfo {pages}
  {023513} (\bibinfo {year} {2018})},\ \Eprint
  {http://arxiv.org/abs/1804.05921} {arXiv:1804.05921 [astro-ph.CO]}
  \BibitemShut {NoStop}%
\bibitem [{\citenamefont {Colpi}\ \emph {et~al.}(1986)\citenamefont {Colpi},
  \citenamefont {Shapiro},\ and\ \citenamefont {Wasserman}}]{Colpi:1986ye}%
  \BibitemOpen
  \bibfield  {author} {\bibinfo {author} {\bibfnamefont {M.}~\bibnamefont
  {Colpi}}, \bibinfo {author} {\bibfnamefont {S.~L.}\ \bibnamefont {Shapiro}},
  \ and\ \bibinfo {author} {\bibfnamefont {I.}~\bibnamefont {Wasserman}},\
  }\href {\doibase 10.1103/PhysRevLett.57.2485} {\bibfield  {journal} {\bibinfo
   {journal} {Phys. Rev. Lett.}\ }\textbf {\bibinfo {volume} {57}},\ \bibinfo
  {pages} {2485} (\bibinfo {year} {1986})}\BibitemShut {NoStop}%
\bibitem [{\citenamefont {Schunck}\ and\ \citenamefont
  {Mielke}(2003)}]{Schunck:2003kk}%
  \BibitemOpen
  \bibfield  {author} {\bibinfo {author} {\bibfnamefont {F.~E.}\ \bibnamefont
  {Schunck}}\ and\ \bibinfo {author} {\bibfnamefont {E.~W.}\ \bibnamefont
  {Mielke}},\ }\href {\doibase 10.1088/0264-9381/20/20/201} {\bibfield
  {journal} {\bibinfo  {journal} {Class. Quant. Grav.}\ }\textbf {\bibinfo
  {volume} {20}},\ \bibinfo {pages} {R301} (\bibinfo {year} {2003})},\ \Eprint
  {http://arxiv.org/abs/0801.0307} {arXiv:0801.0307 [astro-ph]} \BibitemShut
  {NoStop}%
\bibitem [{\citenamefont {Jetzer}(1992)}]{Jetzer:1991jr}%
  \BibitemOpen
  \bibfield  {author} {\bibinfo {author} {\bibfnamefont {P.}~\bibnamefont
  {Jetzer}},\ }\href {\doibase 10.1016/0370-1573(92)90123-H} {\bibfield
  {journal} {\bibinfo  {journal} {Phys. Rept.}\ }\textbf {\bibinfo {volume}
  {220}},\ \bibinfo {pages} {163} (\bibinfo {year} {1992})}\BibitemShut
  {NoStop}%
\bibitem [{\citenamefont {Croon}\ \emph
  {et~al.}(2018{\natexlab{c}})\citenamefont {Croon}, \citenamefont {Gleiser},
  \citenamefont {Mohapatra},\ and\ \citenamefont {Sun}}]{Croon:2018ftb}%
  \BibitemOpen
  \bibfield  {author} {\bibinfo {author} {\bibfnamefont {D.}~\bibnamefont
  {Croon}}, \bibinfo {author} {\bibfnamefont {M.}~\bibnamefont {Gleiser}},
  \bibinfo {author} {\bibfnamefont {S.}~\bibnamefont {Mohapatra}}, \ and\
  \bibinfo {author} {\bibfnamefont {C.}~\bibnamefont {Sun}},\ }\href {\doibase
  10.1016/j.physletb.2018.03.055} {\bibfield  {journal} {\bibinfo  {journal}
  {Phys. Lett.}\ }\textbf {\bibinfo {volume} {B783}},\ \bibinfo {pages} {158}
  (\bibinfo {year} {2018}{\natexlab{c}})},\ \Eprint
  {http://arxiv.org/abs/1802.08259} {arXiv:1802.08259 [hep-ph]} \BibitemShut
  {NoStop}%
\bibitem [{\citenamefont {Helfer}\ \emph {et~al.}(2018)\citenamefont {Helfer},
  \citenamefont {Lim}, \citenamefont {Garcia},\ and\ \citenamefont
  {Amin}}]{Helfer:2018vtq}%
  \BibitemOpen
  \bibfield  {author} {\bibinfo {author} {\bibfnamefont {T.}~\bibnamefont
  {Helfer}}, \bibinfo {author} {\bibfnamefont {E.~A.}\ \bibnamefont {Lim}},
  \bibinfo {author} {\bibfnamefont {M.~A.~G.}\ \bibnamefont {Garcia}}, \ and\
  \bibinfo {author} {\bibfnamefont {M.~A.}\ \bibnamefont {Amin}},\ }\href@noop
  {} {\  (\bibinfo {year} {2018})},\ \Eprint {http://arxiv.org/abs/1802.06733}
  {arXiv:1802.06733 [gr-qc]} \BibitemShut {NoStop}%
\bibitem [{\citenamefont {Widdicombe}\ \emph {et~al.}(2018)\citenamefont
  {Widdicombe}, \citenamefont {Helfer}, \citenamefont {Marsh},\ and\
  \citenamefont {Lim}}]{Widdicombe:2018oeo}%
  \BibitemOpen
  \bibfield  {author} {\bibinfo {author} {\bibfnamefont {J.~Y.}\ \bibnamefont
  {Widdicombe}}, \bibinfo {author} {\bibfnamefont {T.}~\bibnamefont {Helfer}},
  \bibinfo {author} {\bibfnamefont {D.~J.~E.}\ \bibnamefont {Marsh}}, \ and\
  \bibinfo {author} {\bibfnamefont {E.~A.}\ \bibnamefont {Lim}},\ }\href@noop
  {} {\  (\bibinfo {year} {2018})},\ \Eprint {http://arxiv.org/abs/1806.09367}
  {arXiv:1806.09367 [astro-ph.CO]} \BibitemShut {NoStop}%
\bibitem [{\citenamefont {Bezares}\ and\ \citenamefont
  {Palenzuela}(2018)}]{Bezares:2018qwa}%
  \BibitemOpen
  \bibfield  {author} {\bibinfo {author} {\bibfnamefont {M.}~\bibnamefont
  {Bezares}}\ and\ \bibinfo {author} {\bibfnamefont {C.}~\bibnamefont
  {Palenzuela}},\ }\href@noop {} {\  (\bibinfo {year} {2018})},\ \Eprint
  {http://arxiv.org/abs/1808.10732} {arXiv:1808.10732 [gr-qc]} \BibitemShut
  {NoStop}%
\bibitem [{\citenamefont {Chavanis}(2011)}]{Chavanis:2011zi}%
  \BibitemOpen
  \bibfield  {author} {\bibinfo {author} {\bibfnamefont {P.-H.}\ \bibnamefont
  {Chavanis}},\ }\href {\doibase 10.1103/PhysRevD.84.043531} {\bibfield
  {journal} {\bibinfo  {journal} {Phys. Rev.}\ }\textbf {\bibinfo {volume}
  {D84}},\ \bibinfo {pages} {043531} (\bibinfo {year} {2011})},\ \Eprint
  {http://arxiv.org/abs/1103.2050} {arXiv:1103.2050 [astro-ph.CO]} \BibitemShut
  {NoStop}%
\bibitem [{\citenamefont {Chavanis}\ and\ \citenamefont
  {Delfini}(2011)}]{Chavanis:2011zm}%
  \BibitemOpen
  \bibfield  {author} {\bibinfo {author} {\bibfnamefont {P.~H.}\ \bibnamefont
  {Chavanis}}\ and\ \bibinfo {author} {\bibfnamefont {L.}~\bibnamefont
  {Delfini}},\ }\href {\doibase 10.1103/PhysRevD.84.043532} {\bibfield
  {journal} {\bibinfo  {journal} {Phys. Rev.}\ }\textbf {\bibinfo {volume}
  {D84}},\ \bibinfo {pages} {043532} (\bibinfo {year} {2011})},\ \Eprint
  {http://arxiv.org/abs/1103.2054} {arXiv:1103.2054 [astro-ph.CO]} \BibitemShut
  {NoStop}%
\bibitem [{\citenamefont {Chavanis}\ and\ \citenamefont
  {Harko}(2012)}]{Chavanis:2011cz}%
  \BibitemOpen
  \bibfield  {author} {\bibinfo {author} {\bibfnamefont {P.-H.}\ \bibnamefont
  {Chavanis}}\ and\ \bibinfo {author} {\bibfnamefont {T.}~\bibnamefont
  {Harko}},\ }\href {\doibase 10.1103/PhysRevD.86.064011} {\bibfield  {journal}
  {\bibinfo  {journal} {Phys. Rev.}\ }\textbf {\bibinfo {volume} {D86}},\
  \bibinfo {pages} {064011} (\bibinfo {year} {2012})},\ \Eprint
  {http://arxiv.org/abs/1108.3986} {arXiv:1108.3986 [astro-ph.SR]} \BibitemShut
  {NoStop}%
\bibitem [{\citenamefont {Fan}(2016)}]{Fan:2016rda}%
  \BibitemOpen
  \bibfield  {author} {\bibinfo {author} {\bibfnamefont {J.}~\bibnamefont
  {Fan}},\ }\href {\doibase 10.1016/j.dark.2016.10.005} {\bibfield  {journal}
  {\bibinfo  {journal} {Phys. Dark Univ.}\ }\textbf {\bibinfo {volume} {14}},\
  \bibinfo {pages} {84} (\bibinfo {year} {2016})},\ \Eprint
  {http://arxiv.org/abs/1603.06580} {arXiv:1603.06580 [hep-ph]} \BibitemShut
  {NoStop}%
\bibitem [{\citenamefont {Eby}\ \emph {et~al.}(2016{\natexlab{a}})\citenamefont
  {Eby}, \citenamefont {Kouvaris}, \citenamefont {Nielsen},\ and\ \citenamefont
  {Wijewardhana}}]{Eby:2015hsq}%
  \BibitemOpen
  \bibfield  {author} {\bibinfo {author} {\bibfnamefont {J.}~\bibnamefont
  {Eby}}, \bibinfo {author} {\bibfnamefont {C.}~\bibnamefont {Kouvaris}},
  \bibinfo {author} {\bibfnamefont {N.~G.}\ \bibnamefont {Nielsen}}, \ and\
  \bibinfo {author} {\bibfnamefont {L.~C.~R.}\ \bibnamefont {Wijewardhana}},\
  }\href {\doibase 10.1007/JHEP02(2016)028} {\bibfield  {journal} {\bibinfo
  {journal} {JHEP}\ }\textbf {\bibinfo {volume} {02}},\ \bibinfo {pages} {028}
  (\bibinfo {year} {2016}{\natexlab{a}})},\ \Eprint
  {http://arxiv.org/abs/1511.04474} {arXiv:1511.04474 [hep-ph]} \BibitemShut
  {NoStop}%
\bibitem [{\citenamefont {Eby}\ \emph {et~al.}(2018{\natexlab{b}})\citenamefont
  {Eby}, \citenamefont {Leembruggen}, \citenamefont {Street}, \citenamefont
  {Suranyi},\ and\ \citenamefont {Wijewardhana}}]{Eby:2018dat}%
  \BibitemOpen
  \bibfield  {author} {\bibinfo {author} {\bibfnamefont {J.}~\bibnamefont
  {Eby}}, \bibinfo {author} {\bibfnamefont {M.}~\bibnamefont {Leembruggen}},
  \bibinfo {author} {\bibfnamefont {L.}~\bibnamefont {Street}}, \bibinfo
  {author} {\bibfnamefont {P.}~\bibnamefont {Suranyi}}, \ and\ \bibinfo
  {author} {\bibfnamefont {L.~C.~R.}\ \bibnamefont {Wijewardhana}},\
  }\href@noop {} {\  (\bibinfo {year} {2018}{\natexlab{b}})},\ \Eprint
  {http://arxiv.org/abs/1809.08598} {arXiv:1809.08598 [hep-ph]} \BibitemShut
  {NoStop}%
\bibitem [{\citenamefont {Gleiser}(1988)}]{Gleiser:1988rq}%
  \BibitemOpen
  \bibfield  {author} {\bibinfo {author} {\bibfnamefont {M.}~\bibnamefont
  {Gleiser}},\ }\href {\doibase 10.1103/PhysRevD.38.2376,
  10.1103/PhysRevD.39.1257} {\bibfield  {journal} {\bibinfo  {journal} {Phys.
  Rev.}\ }\textbf {\bibinfo {volume} {D38}},\ \bibinfo {pages} {2376} (\bibinfo
  {year} {1988})},\ \bibinfo {note} {[Erratum: Phys.
  Rev.D39,no.4,1257(1989)]}\BibitemShut {NoStop}%
\bibitem [{\citenamefont {Jetzer}(1989)}]{Jetzer:1988vr}%
  \BibitemOpen
  \bibfield  {author} {\bibinfo {author} {\bibfnamefont {P.}~\bibnamefont
  {Jetzer}},\ }\href {\doibase 10.1016/0550-3213(89)90038-2} {\bibfield
  {journal} {\bibinfo  {journal} {Nucl. Phys.}\ }\textbf {\bibinfo {volume}
  {B316}},\ \bibinfo {pages} {411} (\bibinfo {year} {1989})}\BibitemShut
  {NoStop}%
\bibitem [{\citenamefont {Harko}(2014)}]{Harko:2014vya}%
  \BibitemOpen
  \bibfield  {author} {\bibinfo {author} {\bibfnamefont {T.}~\bibnamefont
  {Harko}},\ }\href {\doibase 10.1103/PhysRevD.89.084040} {\bibfield  {journal}
  {\bibinfo  {journal} {Phys. Rev.}\ }\textbf {\bibinfo {volume} {D89}},\
  \bibinfo {pages} {084040} (\bibinfo {year} {2014})},\ \Eprint
  {http://arxiv.org/abs/1403.3358} {arXiv:1403.3358 [gr-qc]} \BibitemShut
  {NoStop}%
\bibitem [{\citenamefont {Eby}\ \emph {et~al.}(2016{\natexlab{b}})\citenamefont
  {Eby}, \citenamefont {Leembruggen}, \citenamefont {Suranyi},\ and\
  \citenamefont {Wijewardhana}}]{Eby:2016cnq}%
  \BibitemOpen
  \bibfield  {author} {\bibinfo {author} {\bibfnamefont {J.}~\bibnamefont
  {Eby}}, \bibinfo {author} {\bibfnamefont {M.}~\bibnamefont {Leembruggen}},
  \bibinfo {author} {\bibfnamefont {P.}~\bibnamefont {Suranyi}}, \ and\
  \bibinfo {author} {\bibfnamefont {L.~C.~R.}\ \bibnamefont {Wijewardhana}},\
  }\href {\doibase 10.1007/JHEP12(2016)066} {\bibfield  {journal} {\bibinfo
  {journal} {JHEP}\ }\textbf {\bibinfo {volume} {12}},\ \bibinfo {pages} {066}
  (\bibinfo {year} {2016}{\natexlab{b}})},\ \Eprint
  {http://arxiv.org/abs/1608.06911} {arXiv:1608.06911 [astro-ph.CO]}
  \BibitemShut {NoStop}%
\bibitem [{\citenamefont {Levkov}\ \emph {et~al.}(2017)\citenamefont {Levkov},
  \citenamefont {Panin},\ and\ \citenamefont {Tkachev}}]{Levkov:2016rkk}%
  \BibitemOpen
  \bibfield  {author} {\bibinfo {author} {\bibfnamefont {D.~G.}\ \bibnamefont
  {Levkov}}, \bibinfo {author} {\bibfnamefont {A.~G.}\ \bibnamefont {Panin}}, \
  and\ \bibinfo {author} {\bibfnamefont {I.~I.}\ \bibnamefont {Tkachev}},\
  }\href {\doibase 10.1103/PhysRevLett.118.011301} {\bibfield  {journal}
  {\bibinfo  {journal} {Phys. Rev. Lett.}\ }\textbf {\bibinfo {volume} {118}},\
  \bibinfo {pages} {011301} (\bibinfo {year} {2017})},\ \Eprint
  {http://arxiv.org/abs/1609.03611} {arXiv:1609.03611 [astro-ph.CO]}
  \BibitemShut {NoStop}%
\bibitem [{\citenamefont {Eby}\ \emph {et~al.}(2018{\natexlab{c}})\citenamefont
  {Eby}, \citenamefont {Leembruggen}, \citenamefont {Suranyi},\ and\
  \citenamefont {Wijewardhana}}]{Eby:2018zlv}%
  \BibitemOpen
  \bibfield  {author} {\bibinfo {author} {\bibfnamefont {J.}~\bibnamefont
  {Eby}}, \bibinfo {author} {\bibfnamefont {M.}~\bibnamefont {Leembruggen}},
  \bibinfo {author} {\bibfnamefont {P.}~\bibnamefont {Suranyi}}, \ and\
  \bibinfo {author} {\bibfnamefont {L.~C.~R.}\ \bibnamefont {Wijewardhana}},\
  }\href@noop {} {\  (\bibinfo {year} {2018}{\natexlab{c}})},\ \Eprint
  {http://arxiv.org/abs/1805.12147} {arXiv:1805.12147 [astro-ph.CO]}
  \BibitemShut {NoStop}%
\bibitem [{\citenamefont {Khan}\ \emph {et~al.}(2016)\citenamefont {Khan},
  \citenamefont {Husa}, \citenamefont {Hannam}, \citenamefont {Ohme},
  \citenamefont {Pürrer}, \citenamefont {Jiménez~Forteza},\ and\
  \citenamefont {Bohé}}]{Khan:2015jqa}%
  \BibitemOpen
  \bibfield  {author} {\bibinfo {author} {\bibfnamefont {S.}~\bibnamefont
  {Khan}}, \bibinfo {author} {\bibfnamefont {S.}~\bibnamefont {Husa}}, \bibinfo
  {author} {\bibfnamefont {M.}~\bibnamefont {Hannam}}, \bibinfo {author}
  {\bibfnamefont {F.}~\bibnamefont {Ohme}}, \bibinfo {author} {\bibfnamefont
  {M.}~\bibnamefont {Pürrer}}, \bibinfo {author} {\bibfnamefont
  {X.}~\bibnamefont {Jiménez~Forteza}}, \ and\ \bibinfo {author}
  {\bibfnamefont {A.}~\bibnamefont {Bohé}},\ }\href {\doibase
  10.1103/PhysRevD.93.044007} {\bibfield  {journal} {\bibinfo  {journal} {Phys.
  Rev.}\ }\textbf {\bibinfo {volume} {D93}},\ \bibinfo {pages} {044007}
  (\bibinfo {year} {2016})},\ \Eprint {http://arxiv.org/abs/1508.07253}
  {arXiv:1508.07253 [gr-qc]} \BibitemShut {NoStop}%
\bibitem [{\citenamefont {Barsotti}\ \emph {et~al.}()\citenamefont {Barsotti},
  \citenamefont {Fritschel}, \citenamefont {Evans},\ and\ \citenamefont
  {Gras}}]{LIGONoise:2018}%
  \BibitemOpen
  \bibfield  {author} {\bibinfo {author} {\bibfnamefont {L.}~\bibnamefont
  {Barsotti}}, \bibinfo {author} {\bibfnamefont {P.}~\bibnamefont {Fritschel}},
  \bibinfo {author} {\bibfnamefont {M.}~\bibnamefont {Evans}}, \ and\ \bibinfo
  {author} {\bibfnamefont {S.}~\bibnamefont {Gras}},\ }\href@noop {} {\
  }\bibinfo {note} {LIGO Document T1800044-v5,
  https://dcc.ligo.org/LIGO-T1800044/public}\BibitemShut {NoStop}%
\bibitem [{\citenamefont {Dominik}\ \emph {et~al.}(2015)\citenamefont
  {Dominik}, \citenamefont {Berti}, \citenamefont {O'Shaughnessy},
  \citenamefont {Mandel}, \citenamefont {Belczynski}, \citenamefont {Fryer},
  \citenamefont {Holz}, \citenamefont {Bulik},\ and\ \citenamefont
  {Pannarale}}]{Dominik:2014yma}%
  \BibitemOpen
  \bibfield  {author} {\bibinfo {author} {\bibfnamefont {M.}~\bibnamefont
  {Dominik}}, \bibinfo {author} {\bibfnamefont {E.}~\bibnamefont {Berti}},
  \bibinfo {author} {\bibfnamefont {R.}~\bibnamefont {O'Shaughnessy}}, \bibinfo
  {author} {\bibfnamefont {I.}~\bibnamefont {Mandel}}, \bibinfo {author}
  {\bibfnamefont {K.}~\bibnamefont {Belczynski}}, \bibinfo {author}
  {\bibfnamefont {C.}~\bibnamefont {Fryer}}, \bibinfo {author} {\bibfnamefont
  {D.~E.}\ \bibnamefont {Holz}}, \bibinfo {author} {\bibfnamefont
  {T.}~\bibnamefont {Bulik}}, \ and\ \bibinfo {author} {\bibfnamefont
  {F.}~\bibnamefont {Pannarale}},\ }\href {\doibase
  10.1088/0004-637X/806/2/263} {\bibfield  {journal} {\bibinfo  {journal}
  {Astrophys. J.}\ }\textbf {\bibinfo {volume} {806}},\ \bibinfo {pages} {263}
  (\bibinfo {year} {2015})},\ \Eprint {http://arxiv.org/abs/1405.7016}
  {arXiv:1405.7016 [astro-ph.HE]} \BibitemShut {NoStop}%
\bibitem [{\citenamefont {Feynman}(1996)}]{Feynman:1996kb}%
  \BibitemOpen
  \bibfield  {author} {\bibinfo {author} {\bibfnamefont {R.~P.}\ \bibnamefont
  {Feynman}},\ }\href@noop {} {\emph {\bibinfo {title} {{Feynman lectures on
  gravitation}}}},\ edited by\ \bibinfo {editor} {\bibfnamefont {F.~B.}\
  \bibnamefont {Morinigo}}, \bibinfo {editor} {\bibfnamefont {W.~G.}\
  \bibnamefont {Wagner}}, \ and\ \bibinfo {editor} {\bibfnamefont
  {B.}~\bibnamefont {Hatfield}}\ (\bibinfo {year} {1996})\BibitemShut {NoStop}%
\bibitem [{\citenamefont {De~Hoog}\ and\ \citenamefont
  {Weiss}(1976)}]{de1976difference}%
  \BibitemOpen
  \bibfield  {author} {\bibinfo {author} {\bibfnamefont {F.~R.}\ \bibnamefont
  {De~Hoog}}\ and\ \bibinfo {author} {\bibfnamefont {R.}~\bibnamefont
  {Weiss}},\ }\href@noop {} {\bibfield  {journal} {\bibinfo  {journal} {SIAM
  Journal on Numerical Analysis}\ }\textbf {\bibinfo {volume} {13}},\ \bibinfo
  {pages} {775} (\bibinfo {year} {1976})}\BibitemShut {NoStop}%
\bibitem [{\citenamefont {Weinm{\"u}ller}(1984)}]{weinmuller1984boundary}%
  \BibitemOpen
  \bibfield  {author} {\bibinfo {author} {\bibfnamefont {E.}~\bibnamefont
  {Weinm{\"u}ller}},\ }\href@noop {} {\bibfield  {journal} {\bibinfo  {journal}
  {SIAM journal on mathematical analysis}\ }\textbf {\bibinfo {volume} {15}},\
  \bibinfo {pages} {287} (\bibinfo {year} {1984})}\BibitemShut {NoStop}%
\bibitem [{\citenamefont {Weinm{\"u}ller}(1986)}]{weinmuller1986collocation}%
  \BibitemOpen
  \bibfield  {author} {\bibinfo {author} {\bibfnamefont {E.}~\bibnamefont
  {Weinm{\"u}ller}},\ }\href@noop {} {\bibfield  {journal} {\bibinfo  {journal}
  {SIAM journal on numerical analysis}\ }\textbf {\bibinfo {volume} {23}},\
  \bibinfo {pages} {1062} (\bibinfo {year} {1986})}\BibitemShut {NoStop}%
\bibitem [{\citenamefont {Burkotov{\'a}}\ \emph {et~al.}(2016)\citenamefont
  {Burkotov{\'a}}, \citenamefont {Hubner}, \citenamefont {Rachunkov{\'a}},\
  and\ \citenamefont {Weinm{\"u}ller}}]{burkotova2016asymptotic}%
  \BibitemOpen
  \bibfield  {author} {\bibinfo {author} {\bibfnamefont {J.}~\bibnamefont
  {Burkotov{\'a}}}, \bibinfo {author} {\bibfnamefont {M.}~\bibnamefont
  {Hubner}}, \bibinfo {author} {\bibfnamefont {I.}~\bibnamefont
  {Rachunkov{\'a}}}, \ and\ \bibinfo {author} {\bibfnamefont {E.~B.}\
  \bibnamefont {Weinm{\"u}ller}},\ }\href@noop {} {\bibfield  {journal}
  {\bibinfo  {journal} {Applied Mathematics and Computation}\ }\textbf
  {\bibinfo {volume} {274}},\ \bibinfo {pages} {65} (\bibinfo {year}
  {2016})}\BibitemShut {NoStop}%
\bibitem [{\citenamefont {Burkotov{\'a}}\ \emph {et~al.}(2018)\citenamefont
  {Burkotov{\'a}}, \citenamefont {Rachunkov{\'a}}, \citenamefont
  {Stan{\v{e}}k}, \citenamefont {Weinm{\"u}ller},\ and\ \citenamefont
  {Wurm}}]{burkotova2018nonlinear}%
  \BibitemOpen
  \bibfield  {author} {\bibinfo {author} {\bibfnamefont {J.}~\bibnamefont
  {Burkotov{\'a}}}, \bibinfo {author} {\bibfnamefont {I.}~\bibnamefont
  {Rachunkov{\'a}}}, \bibinfo {author} {\bibfnamefont {S.}~\bibnamefont
  {Stan{\v{e}}k}}, \bibinfo {author} {\bibfnamefont {E.~B.}\ \bibnamefont
  {Weinm{\"u}ller}}, \ and\ \bibinfo {author} {\bibfnamefont {S.}~\bibnamefont
  {Wurm}},\ }\href@noop {} {\bibfield  {journal} {\bibinfo  {journal} {Applied
  Numerical Mathematics}\ }\textbf {\bibinfo {volume} {130}},\ \bibinfo {pages}
  {23} (\bibinfo {year} {2018})}\BibitemShut {NoStop}%
\bibitem [{\citenamefont {Koch}\ and\ \citenamefont
  {Weinm{\"u}ller}(2003)}]{koch2003convergence}%
  \BibitemOpen
  \bibfield  {author} {\bibinfo {author} {\bibfnamefont {O.}~\bibnamefont
  {Koch}}\ and\ \bibinfo {author} {\bibfnamefont {E.}~\bibnamefont
  {Weinm{\"u}ller}},\ }\href@noop {} {\bibfield  {journal} {\bibinfo  {journal}
  {Mathematics of computation}\ }\textbf {\bibinfo {volume} {72}},\ \bibinfo
  {pages} {289} (\bibinfo {year} {2003})}\BibitemShut {NoStop}%
\bibitem [{\citenamefont {Weinm{\"u}ller}()}]{SBVPshooting}%
  \BibitemOpen
  \bibfield  {author} {\bibinfo {author} {\bibfnamefont {E.}~\bibnamefont
  {Weinm{\"u}ller}},\ }\href@noop {} {\ }\bibinfo {note} {Private
  communication.}\BibitemShut {Stop}%
\bibitem [{\citenamefont {Kitzhofer}\ \emph {et~al.}(2009)\citenamefont
  {Kitzhofer}, \citenamefont {Koch}, \citenamefont {Pulverer}, \citenamefont
  {Simon},\ and\ \citenamefont {Weinm{\"u}ller}}]{kitzhofer2009bvpsuite}%
  \BibitemOpen
  \bibfield  {author} {\bibinfo {author} {\bibfnamefont {G.}~\bibnamefont
  {Kitzhofer}}, \bibinfo {author} {\bibfnamefont {O.}~\bibnamefont {Koch}},
  \bibinfo {author} {\bibfnamefont {G.}~\bibnamefont {Pulverer}}, \bibinfo
  {author} {\bibfnamefont {C.}~\bibnamefont {Simon}}, \ and\ \bibinfo {author}
  {\bibfnamefont {E.~B.}\ \bibnamefont {Weinm{\"u}ller}},\ }\href@noop {}
  {\bibfield  {journal} {\bibinfo  {journal} {ASC Report No}\ }\textbf
  {\bibinfo {volume} {35}} (\bibinfo {year} {2009})}\BibitemShut {NoStop}%
\bibitem [{\citenamefont {Kitzhofer}\ \emph {et~al.}(2010)\citenamefont
  {Kitzhofer}, \citenamefont {Koch}, \citenamefont {Pulverer}, \citenamefont
  {Simon},\ and\ \citenamefont {Weinm{\"u}ller}}]{kitzhofer2010new}%
  \BibitemOpen
  \bibfield  {author} {\bibinfo {author} {\bibfnamefont {G.}~\bibnamefont
  {Kitzhofer}}, \bibinfo {author} {\bibfnamefont {O.}~\bibnamefont {Koch}},
  \bibinfo {author} {\bibfnamefont {G.}~\bibnamefont {Pulverer}}, \bibinfo
  {author} {\bibfnamefont {C.}~\bibnamefont {Simon}}, \ and\ \bibinfo {author}
  {\bibfnamefont {E.}~\bibnamefont {Weinm{\"u}ller}},\ }\href@noop {}
  {\bibfield  {journal} {\bibinfo  {journal} {J. Numer. Anal. Indust. Appl.
  Math}\ }\textbf {\bibinfo {volume} {5}},\ \bibinfo {pages} {113} (\bibinfo
  {year} {2010})}\BibitemShut {NoStop}%
\bibitem [{\citenamefont {Arnowitt}\ \emph {et~al.}(2008)\citenamefont
  {Arnowitt}, \citenamefont {Deser},\ and\ \citenamefont
  {Misner}}]{Arnowitt:1962hi}%
  \BibitemOpen
  \bibfield  {author} {\bibinfo {author} {\bibfnamefont {R.~L.}\ \bibnamefont
  {Arnowitt}}, \bibinfo {author} {\bibfnamefont {S.}~\bibnamefont {Deser}}, \
  and\ \bibinfo {author} {\bibfnamefont {C.~W.}\ \bibnamefont {Misner}},\
  }\href {\doibase 10.1007/s10714-008-0661-1} {\bibfield  {journal} {\bibinfo
  {journal} {Gen. Rel. Grav.}\ }\textbf {\bibinfo {volume} {40}},\ \bibinfo
  {pages} {1997} (\bibinfo {year} {2008})},\ \Eprint
  {http://arxiv.org/abs/gr-qc/0405109} {arXiv:gr-qc/0405109 [gr-qc]}
  \BibitemShut {NoStop}%
\bibitem [{\citenamefont {Carroll}(2004)}]{Carroll:2004st}%
  \BibitemOpen
  \bibfield  {author} {\bibinfo {author} {\bibfnamefont {S.~M.}\ \bibnamefont
  {Carroll}},\ }\href
  {http://www.slac.stanford.edu/spires/find/books/www?cl=QC6:C37:2004} {\emph
  {\bibinfo {title} {{Spacetime and geometry: An introduction to general
  relativity}}}}\ (\bibinfo {year} {2004})\BibitemShut {NoStop}%
\bibitem [{\citenamefont {Soni}\ and\ \citenamefont
  {Zhang}(2016)}]{Soni:2016gzf}%
  \BibitemOpen
  \bibfield  {author} {\bibinfo {author} {\bibfnamefont {A.}~\bibnamefont
  {Soni}}\ and\ \bibinfo {author} {\bibfnamefont {Y.}~\bibnamefont {Zhang}},\
  }\href {\doibase 10.1103/PhysRevD.93.115025} {\bibfield  {journal} {\bibinfo
  {journal} {Phys. Rev.}\ }\textbf {\bibinfo {volume} {D93}},\ \bibinfo {pages}
  {115025} (\bibinfo {year} {2016})},\ \Eprint
  {http://arxiv.org/abs/1602.00714} {arXiv:1602.00714 [hep-ph]} \BibitemShut
  {NoStop}%
\bibitem [{\citenamefont {Eby}\ \emph {et~al.}(2016{\natexlab{c}})\citenamefont
  {Eby}, \citenamefont {Suranyi},\ and\ \citenamefont
  {Wijewardhana}}]{Eby:2015hyx}%
  \BibitemOpen
  \bibfield  {author} {\bibinfo {author} {\bibfnamefont {J.}~\bibnamefont
  {Eby}}, \bibinfo {author} {\bibfnamefont {P.}~\bibnamefont {Suranyi}}, \ and\
  \bibinfo {author} {\bibfnamefont {L.~C.~R.}\ \bibnamefont {Wijewardhana}},\
  }\href {\doibase 10.1142/S0217732316500905} {\bibfield  {journal} {\bibinfo
  {journal} {Mod. Phys. Lett.}\ }\textbf {\bibinfo {volume} {A31}},\ \bibinfo
  {pages} {1650090} (\bibinfo {year} {2016}{\natexlab{c}})},\ \Eprint
  {http://arxiv.org/abs/1512.01709} {arXiv:1512.01709 [hep-ph]} \BibitemShut
  {NoStop}%
\bibitem [{\citenamefont {Helfer}\ \emph {et~al.}(2017)\citenamefont {Helfer},
  \citenamefont {Marsh}, \citenamefont {Clough}, \citenamefont {Fairbairn},
  \citenamefont {Lim},\ and\ \citenamefont {Becerril}}]{Helfer:2016ljl}%
  \BibitemOpen
  \bibfield  {author} {\bibinfo {author} {\bibfnamefont {T.}~\bibnamefont
  {Helfer}}, \bibinfo {author} {\bibfnamefont {D.~J.~E.}\ \bibnamefont
  {Marsh}}, \bibinfo {author} {\bibfnamefont {K.}~\bibnamefont {Clough}},
  \bibinfo {author} {\bibfnamefont {M.}~\bibnamefont {Fairbairn}}, \bibinfo
  {author} {\bibfnamefont {E.~A.}\ \bibnamefont {Lim}}, \ and\ \bibinfo
  {author} {\bibfnamefont {R.}~\bibnamefont {Becerril}},\ }\href {\doibase
  10.1088/1475-7516/2017/03/055} {\bibfield  {journal} {\bibinfo  {journal}
  {JCAP}\ }\textbf {\bibinfo {volume} {1703}},\ \bibinfo {pages} {055}
  (\bibinfo {year} {2017})},\ \Eprint {http://arxiv.org/abs/1609.04724}
  {arXiv:1609.04724 [astro-ph.CO]} \BibitemShut {NoStop}%
\bibitem [{\citenamefont {Sarkar}\ \emph {et~al.}(2018)\citenamefont {Sarkar},
  \citenamefont {Vaz},\ and\ \citenamefont {Wijewardhana}}]{Sarkar:2017aje}%
  \BibitemOpen
  \bibfield  {author} {\bibinfo {author} {\bibfnamefont {S.}~\bibnamefont
  {Sarkar}}, \bibinfo {author} {\bibfnamefont {C.}~\bibnamefont {Vaz}}, \ and\
  \bibinfo {author} {\bibfnamefont {L.~C.~R.}\ \bibnamefont {Wijewardhana}},\
  }\href {\doibase 10.1103/PhysRevD.97.103022} {\bibfield  {journal} {\bibinfo
  {journal} {Phys. Rev.}\ }\textbf {\bibinfo {volume} {D97}},\ \bibinfo {pages}
  {103022} (\bibinfo {year} {2018})},\ \Eprint
  {http://arxiv.org/abs/1711.01219} {arXiv:1711.01219 [astro-ph.GA]}
  \BibitemShut {NoStop}%
\bibitem [{\citenamefont {Hertzberg}\ and\ \citenamefont
  {Schiappacasse}(2018)}]{Hertzberg:2018lmt}%
  \BibitemOpen
  \bibfield  {author} {\bibinfo {author} {\bibfnamefont {M.~P.}\ \bibnamefont
  {Hertzberg}}\ and\ \bibinfo {author} {\bibfnamefont {E.~D.}\ \bibnamefont
  {Schiappacasse}},\ }\href {\doibase 10.1088/1475-7516/2018/08/028} {\bibfield
   {journal} {\bibinfo  {journal} {JCAP}\ }\textbf {\bibinfo {volume} {1808}},\
  \bibinfo {pages} {028} (\bibinfo {year} {2018})},\ \Eprint
  {http://arxiv.org/abs/1804.07255} {arXiv:1804.07255 [hep-ph]} \BibitemShut
  {NoStop}%
\end{thebibliography}%
\end{document}